\documentclass{aa}

\usepackage{diagbox}
\usepackage{array}
\usepackage{booktabs}
\usepackage{graphicx}
\usepackage{multirow}
\usepackage{xcolor}
\usepackage{soul}
\usepackage{bm}
\usepackage[justification=centering]{caption} 
\usepackage{caption}
\usepackage{subcaption}

\usepackage{float}
\usepackage{stfloats}
\usepackage{graphicx}
\usepackage{natbib}
\usepackage{makecell}

\usepackage{txfonts}
\usepackage{braket}
\def \be {\begin{equation}}
\def \ee {\end{equation}}

\newcommand{\genec}{\textsc{\Large g\large enec\normalsize}}
\newcommand{\syclist}{\textsc{\Large s\large yclist\normalsize}}

\newcommand{\Genec}{\textsc{\Large g\large enec \normalsize}}
\newcommand{\Syclist}{\textsc{\Large s\large yclist \normalsize}}

\newcommand{\mesa}{\textsc{\large mesa \normalsize}}

\defcitealias{pol98}{P98}

\defcitealias{pol98}{P98}

\defcitealias{hur00}{H00}

\defcitealias{hur00}{H00}

\defcitealias{hur02}{H02}

\defcitealias{hur02}{H02}

\defcitealias{sci24}{S24}

\defcitealias{sci24}{S24}

\defcitealias{zah77}{Z77}

\defcitealias{zah77}{Z77}

\usepackage[colorlinks=true, citecolor=blue, linkcolor=blue, urlcolor=blue]{hyperref}

\begin{document}

\title{The IACOB project}
\subtitle{XIX. Revisiting massive-star evolution with empirical TAMS constraints: Updated models, overshoot calibration, and the population of blue supergiants}

\author{Luca Sciarini\inst{1}, Abel de Burgos\inst{2}, 
Sophie Rosu\inst{1},
Sylvia Ekstr\"om\inst{1}, Sergio Simón-Díaz\inst{3,4}, and Cyril Georgy\inst{1}}
\institute{Department of Astronomy, University of Geneva, Chemin Pegasi 51, CH-1290 Versoix, Switzerland\\\email{luca.sciarini@unige.ch}
\and
European Southern Observatory, Alonso de Córdova 3107, Vitacura, Santiago, Chile\and
Instituto de Astrof\'isica de Canarias, Avenida Vía Láctea, E-38205 La Laguna, Tenerife, Spain\and
Universidad de La Laguna, Dpto. Astrof\'isica, E-38206 La Laguna, Tenerife, Spain}

   \date{Received date ... /
Accepted date ...}
\authorrunning{Luca Sciarini et al.}
\titlerunning{Revisiting massive-star evolution with the IACOB sample}

  \abstract
   {Massive stars play a fundamental role in the evolution of the Universe, yet several physical processes governing their evolution remain poorly constrained. Notably, the main-sequence width is known to be sensitive to the convective boundary mixing efficiency. Moreover, discrepancies between single-star evolution predictions and observations highlight the need to account for binary interactions to explain some populations of massive stars.}
   {Our goal is to constrain single-star models using recent observations of massive Galactic stars from the IACOB database. We use the latest proposed empirical location of the terminal age main sequence (TAMS) to calibrate the convective boundary mixing efficiency and use this calibration to test single-star evolution by comparing various model predictions to the observed populations of the IACOB sample.} 
   {We compute detailed single-star models with \genec. We calibrate the convective boundary mixing efficiency, ensuring that the models reproduce the proposed empirical TAMS location. We compute several grids with various overshoot calibrations, angular momentum (AM) transport treatments, initial masses ($M_{\rm ini}=12-40$\,M$_\odot$) and velocities ($\upsilon_{\rm ini}/\upsilon_{\rm crit}=0.0-0.6$). Finally, we generate synthetic populations from the tracks with \Syclist and perform a direct comparison with the observed population.}
   {The calibrated models at slow rotation ($\upsilon_{\rm ini}/\upsilon_{\rm crit}=0.1$) reproduce the empirical TAMS location. We find that a mass-dependent overshoot efficiency is required to fit the observational constraints. Additionally, the overall rotational properties of the observed populations are well reproduced with single-star models, independently of the AM transport assumptions. In particular, models accounting only for hydrodynamical instabilities are successful at reproducing the rotational properties, unlike previous \Genec grids, which we attribute to the choice of wind prescription. Although the empirical TAMS of slow rotators ($\upsilon \sin i<100$\,km/s) is well reproduced, we find that models are unsuccessful at explaining the velocity dependence of the TAMS location observed in the IACOB sample. Finally, we find that single-star models fail at explaining the population of blue supergiants to the right of the TAMS location, as they cross the Hertzsprung gap too fast. The synthetic populations strengthen this result, predicting $\sim 2$\,\% of stars in this region instead of the observed $\sim 15$\,\%.}{}
   

   \maketitle

\section{Introduction}
Massive stars $(M\gtrsim 8\,$M$_\odot)$ are fundamental drivers of cosmic evolution. Through their intense ultraviolet radiation, they ionize their surroundings and drive stellar formation \citep[e.g.,][]{ost06,dal15}. With their powerful stellar winds and supernova ejecta, they enrich their environment in nucleosynthetic products, making them the main drivers of galactic evolution \citep[e.g.,][]{lei99,mat12,nom13}. Stellar models are essential tools for the analysis of astrophysical observations \citep[e.g.,][]{bro11a,eks12,cho16,lim18,cos25}. However, their predictive capacities are limited by uncertainties associated with the poorly constrained parameters they incorporate.

Current and upcoming large-scale spectroscopic surveys, such as the Instituto de Astrof\'isica de Canarias OB stars survey (IACOB; last described in \citealt{sim20}, see also \citealt{sim26a}) or the Binarity at LOw Metallicity campaign \citep[BLOeM;][]{she24} are providing statistically robust observable constraints that new generations of stellar grids should aim to reproduce. For instance, when available, rotational properties help constrain the efficiency of rotational mixing, angular momentum (AM) transport, and/or the amount of angular momentum removal by stellar winds \citep[see, e.g.,][]{hun09,mae09,bro11b,hol22}. Overdensities in observed populations offer clues to their progenitors when they are not predicted by single-star models. The overpopulation of blue supergiants (BSGs), for example, is challenging to explain with single-star evolution. Binary interactions, mergers in particular, have been proposed as channels for forming these objects \citep[see, e.g.,][]{wal64,fit90,egg02,men24}.

Recently, \citet{deb25} proposed an updated empirical terminal age main sequence (TAMS) location derived from the drop of density of stars in the Hertzsprung-Russell diagram (HRD) based on a compilation of spectroscopic observations from the IACOB sample in the luminosity range \mbox{$\log L/\text{L}_\odot=4.30-5.70$}. They showed that widely used grids of stellar tracks -- namely the BONN \citep{bro11a}, GENEVA \citep{eks12}, and MIST models \citep{cho16} -- fail at reproducing the empirical TAMS lines. Their proposed TAMS lines are key observable constraints, and they can be used to improve the physics of stellar models (see also \citealt{cas14}). In particular, the region in the HRD covered by the models during their main-sequence (MS) evolution -- commonly referred to as the MS width -- is known to be very sensitive to the efficiency of convective boundary mixing \citep[CBM; see, e.g.,][]{bar23}.

In this work, we compute grids of stellar models with the GENeva stellar Evolution Code \citep[\genec;][]{egg08} with various AM transport treatments. We performed independent calibrations of the CBM efficiency -- parametrized with the step overshoot parameter $\alpha_{\rm ov}$ -- to fit the empirical TAMS lines by \citet{deb25}. We investigated the impact of rotation on the evolution of the models by extending the grids to a wide range of initial velocities. Finally, we generated synthetic populations from the stellar tracks using the code SYnthetic CLusters, Isochrones, and Stellar Tracks \citep[\syclist;][]{geo14}, which we directly compared with the observed population.

Simultaneous to this work, \citet{lec26} adopted a similar approach to calibrate the overshoot parameter so that \mesa models reproduce the observations by \citet{deb25}, and obtained consistent results. While their calibration relies on a Bayesian inference approach to fit the observed population, ours directly uses the TAMS lines from \citet{deb25}. Our analysis extends further to include comparison of the rotational properties of the models with those of the observed population.

The structure of this paper is as follows: In Sect. \ref{ingredients}, we provide a description of our methods and stellar physics assumptions. In Sect. \ref{stellar_tracks}, we present the stellar tracks and compare them to the IACOB sample. In Sect. \ref{population_comparison}, we compare the synthetic populations with the observed population. In Sect. \ref{discussion}, we discuss the robustness and limitations of our results, and in Sect. \ref{conclusion} we summarize our findings.
\section{Stellar physics ingredients}\label{ingredients}
\subsection{Grids ingredients}
In this study we use \Genec to compute detailed 1D stellar evolution models at solar metallicity ($Z=0.014)$. The adopted stellar physics is the same as in \citet{sci26}, with some differences explicitly listed below.

Earlier studies performed with \Genec adopted the \citet{vin01} winds recipe. Given the increasing evidence that this treatment overestimates the strength of the winds \citep[e.g.,][]{cro06,bjo21,bjo23,bra22,deb24,haw24,krt24}, in this work we use the \citet{bjo23} prescription. Outside the range of its applicability we use the standard wind combination adopted in \Genec models, consisting of the \citet{dej88}, \citet{nug00} and \citet{gra21} prescriptions. Although 15\,M$_\odot$ and 12\,M$_\odot$ models lie at and beyond the lower limit of the \citet{bjo23} prescription, we still apply their recipe instead of that of \citet{dej88}. This choice prevents stars in the range $12-15\,$M$_\odot$ from losing more mass than their more massive counterparts, which we consider unphysical. Regardless, in this mass range the mass loss rates are too low to impact the evolution in the HRD. By default, in \Genec Wolf-Rayet (WR) winds are applied for effective temperature $T_{\rm eff}>10$\,kK and a hydrogen surface mass fraction below 0.4. It was recently shown by \citet{pau25} that the transition to WR winds appears directly related to the proximity to the Eddington limit, parametrized by the Eddington factor $\Gamma_{\rm e}$. They showed that WR stars are all found to have Eddington factors $\log\Gamma_{\rm e}\gtrsim-0.3$. In this work, we added this condition for launching WR winds and then used the prescriptions of \citet{nug00} and \citet{gra21}. We ran the models until they reached $T_{\rm eff}<14$\,kK, which corresponds to the limit of the sample by \citet{deb25}. As such, the red supergiant (RSG) phase is not computed. In Sect. \ref{RSG_disc} we discuss the impact of the strength of RSG mass loss on the post-MS evolution. We find that when the \citet{bea20} prescription is used, models do not evolve to hotter temperatures after the RSG phase. In this case, all the results presented in this study would remain unchanged. Thus, the \citet{bea20} treatment can be seen as our default RSG winds prescription.

We computed models that account for magnetohydrodynamical (hereafter magnetic models) and purely hydrodynamical instabilities (hereafter hydro models). The physics of angular momentum transport and rotational mixing of each type of models is identical to the treatment in \citet{sci26}, in particular the rotational mixing efficiency of the magnetic models (see the calibration in Appendix A of their study). In hydro models the AM transport is computed following an advective-diffusive scheme. In magnetic models, a diffusive approximation is justified since magnetic instabilities (computed following the astroseismic-calibrated version of the Tayler-Spruit dynamo \citealt{tay73,spr02,egg22}) largely dominate the transport \citep{mae05}. For chemical elements, the diffusive approach is followed in both cases. The combined effect of meridional circulation and horizontal turbulence mitigates the advection of chemical elements through homogenization of horizontal layers, reducing the vertical transport to a diffusive process \citep{cha92}.

The physics of convection is the same as in \citet{eks12}, apart from the treatment of overshoot. We use the Schwarzschild criterion and follow the constant step-overshoot recipe. In the whole convective core and the overshooting region, the temperature gradient is the adiabatic one, as in all previous \Genec grids \citep[see, e.g.,][]{mar21}.
\subsection{Overshoot calibration}
Previous grids of \Genec models adopted the value \mbox{$\alpha_{\rm ov}=0.1$}. The value was calibrated using a population of Galactic stars in the mass range $1.35-9\,$M$_\odot$ \citep{eks12}. Observational \citep{cas14,deb25} and theoretical studies \citep{ros20b,ros26,sco21} indicate that this value may be underestimated in the massive star domain ($M\gtrsim 8$\,M$_\odot$). In this work, we compute stellar models in the mass range $12-40$\,M$_\odot$ and systematically calibrate the value of $\alpha_{\rm ov}$ such that the models TAMS line -- defined as in \citet{mar21} with the minimum temperature reached during the MS, that is, the MS hook -- exactly fits the empirical TAMS line derived by \citet{deb25}. This empirical TAMS location was obtained from a sample of $\sim 650$ O stars and B giants and supergiants (full sample) using the drop in density of stars in the HRD as a function of $T_{\rm eff}$. It is parametrized with the linear fit
\begin{equation}
    \log (L/\text{L}_\odot)=0.47 \,T_{\rm eff}-5.42\,[\text{dex}]
\label{TAMS_eq_FS}
\end{equation}
in the luminosity range $\log(L/\text{L}_\odot)=4.3-5.7$, with $T_{\rm eff}$ expressed in kK. In their work, \citet{deb25} also proposed an alternative TAMS line, obtained when only slow rotators (i.e., projected velocities $\upsilon\sin i<100$\,km\,s$^{-1}$) are considered. The TAMS location of the slow rotators is parametrized by the vertical line\footnote{The parametric equation given in \citet{deb25} contains a typo. The correct fit is $T_{\rm eff}=21.45$\,kK (instead of $T_{\rm eff}=22.65$\,kK, which is given in their paper).}
\begin{equation}
T_{\rm eff}=21.45\,\text{kK}.
\label{TAMS_eq_SR}
\end{equation}
We performed an equivalent calibration to reproduce this empirical TAMS line. There are valid reasons to prefer either line. On the one hand, the full sample contains fast rotators ($\upsilon\sin i>100$\,km/s), the majority of which are proposed to be binary interaction products \citep{dem13,hol22,bri23} and therefore are not representative of single-star evolution. Additionally, rotational mixing alters the TAMS location of the models. Thus, it seems a reasonable choice to perform the overshoot calibration with only slow rotators in order to break the degeneracy between the effect of overshoot and rotational mixing when determining the TAMS line of the models. For these reasons, the calibration with only slow rotators is our preferred approach. On the other hand, removing the fast rotators weakens the statistical robustness, as they represent about one third of the full sample. Additionally, the observed rotation rate does not inform us about the history of the star. Some of the stars appearing as slow rotators today may have been fast rotators in the past that slowed down through different processes (e.g., mass loss, tidal spin-down) and are therefore not better suited than fast rotators for calibrating the models. Finally, some of the slow rotators are fast rotators observed pole-on. For these reasons, we decided to perform two independent calibrations of the overshoot efficiency, one with each sample (slow rotators and full sample) and compare the results of the two calibrations.

The bulk of velocities of the population of slow rotators in the IACOB sample close to zero age main sequence (ZAMS) is around $\upsilon\sin i/\upsilon_{\rm crit}=0.075$. This corresponds to initial velocities of about $\upsilon_{\rm ini}/\upsilon_{\rm crit}=0.1$\footnote{Assuming a $\sin i$ distribution with probability distribution $P(i)=\sin i$, with expected value $\braket{\sin i}=\pi/4$ \citep[see, e.g.,][]{gra88}.}  (see \citealt{hol22} and Fig. \ref{init_vel_distr}), which we used as reference value for the calibration of $\alpha_{\rm ov}$. We note that this reference value is substantially lower than that adopted in \citet[$\upsilon_{\rm ini}/\upsilon_{\rm crit}=0.4$]{eks12}. This discrepancy is mainly due to two reasons. First, the value $\upsilon_{\rm ini}/\upsilon_{\rm crit}=0.4$ is meant to reproduce velocities of Galactic stars throughout their MS, whereas here we restricted the sample to stars with surface gravities $\log g_{\rm surf}>3.7$, that is, near ZAMS, for determining initial velocities. Secondly, the value adopted in \citet{eks12} is meant to reproduce the average velocities of the stars and does not consider the bimodality of the velocity distribution \citep[e.g.,][]{con77,how97,duf13,ram13,hol22,deb24}. Given the increasing consensus that a large fraction of fast rotators are binary interaction products, it is more appropriate to perform single-star evolution calibrations with the slow rotator population in our view.

\begin{figure}[h]
\centering
\centerline{\includegraphics[trim=0.3cm 0.3cm 0.3cm 0.2cm, clip=true, width=1.\columnwidth,angle=0]{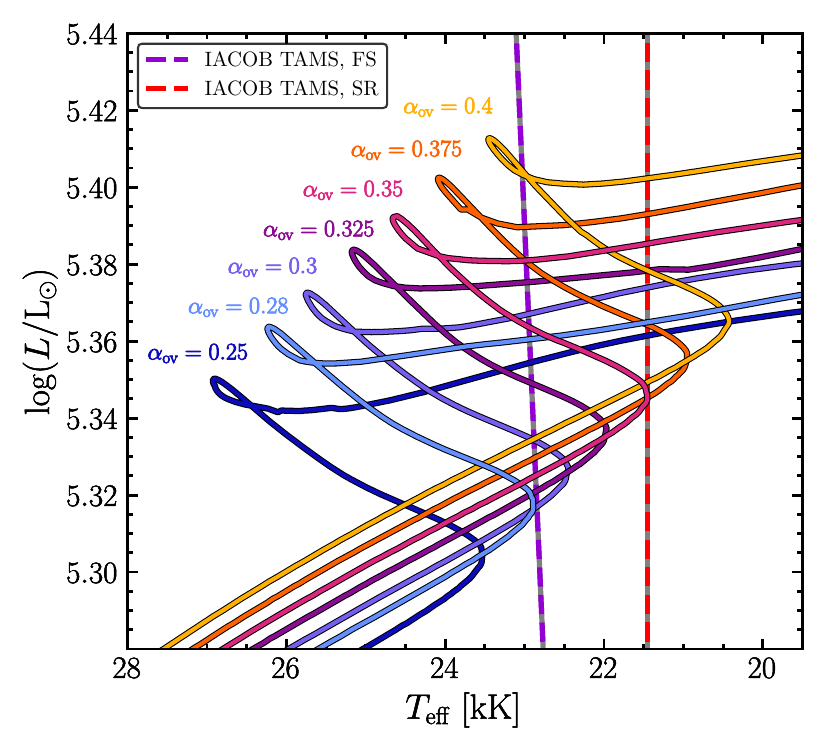}}
\caption{End of the MS and post-MS evolution in the HRD of 25\,M$_\odot$ hydro models with $\upsilon_{\rm ini}/\upsilon_{\rm crit}=0.1$ and overshoot efficiencies in the range \mbox{$\alpha_{\rm ov}=0.25-0.4$} compared to empirical TAMS from \citet{deb25}. Purple-gray line: Full sample TAMS. Red-gray line: Slow rotator TAMS.}
\label{overshoot_calib_25}
\end{figure}

Figure \ref{overshoot_calib_25} illustrates the overshoot efficiency calibration for a 25\,M$_\odot$ hydro model. 
Models were computed with overshoot efficiencies of $\alpha_{\rm ov}=0.25-0.4$. In this range, the theoretical TAMS monotonously shifts to cooler temperatures with increasing values of $\alpha_{\rm ov}$, which guarantees the uniqueness of the calibration. For this choice of initial mass, the theoretical TAMS matches the empirical TAMS with $\alpha_{\rm ov}=0.28$ when considering the full sample, and with $\alpha_{\rm ov}=0.35$ when considering only the slow rotators.

\begin{figure}[h]
\centering
\centerline{\includegraphics[trim=1.9cm 1.cm 1.55cm 1.35cm, clip=true, width=1.\columnwidth,angle=0]{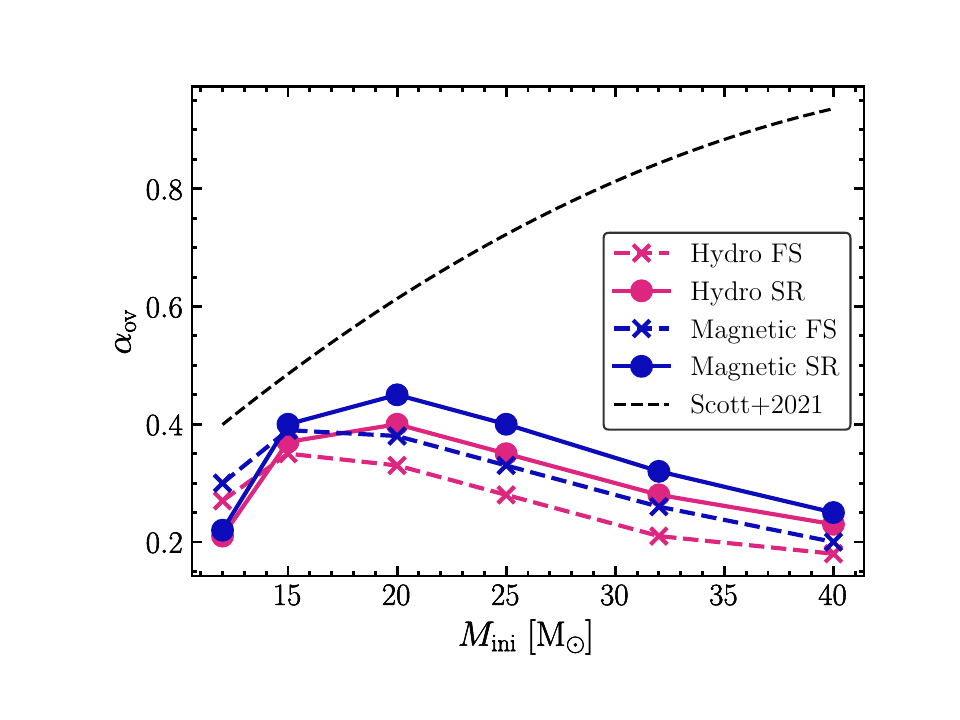}}
\caption{Calibrated values of $\alpha_{\rm ov}$ for stellar models in the mass range $M_{\rm ini}=12-40$\,M$_\odot$ using constraints from the IACOB sample \citep{deb25}. Dashed lines: Calibrated values for the full sample (i.e., Eq. \ref{TAMS_eq_FS}). Solid line: Calibrated values for the slow rotators (i.e., Eq. \ref{TAMS_eq_SR}). Magenta lines: Hydro models. Blue line: Magnetic models. Dashed black line: $\alpha_{\rm ov}$ from \citet{sco21}.}
\label{alpha_ov_calib}
\end{figure}

The calibrated values of $\alpha_{\rm ov}$ obtained with this approach for the full investigated mass range and the two types of AM transport treatments (hydro and magnetic models) are reported in Fig. \ref{alpha_ov_calib} and Table \ref{table_overshoot}. Following this empirical approach, we find that models require overshoot efficiencies in the range \mbox{$\alpha_{\rm ov}\sim0.18-0.45$} depending on the initial mass, AM transport and sample, confirming the need for a value larger than 0.1 in this mass range. We obtain a non-monotonic dependence of $\alpha_{\rm ov}$ with mass. It peaks around $15-20$\,M$_\odot$ and decreases at higher masses, in opposition to the results of \citet{sco21}, who found a monotonic increase of $\alpha_{\rm ov}$ within this mass range (see their Fig. 9). The mass trend appears robust, consistently observed across AM transport and sample choices. We note that hydro models require marginally lower $\alpha_{\rm ov}$ values than magnetic models to fit the empirical TAMS. This is explained by the fact that in magnetic models, rotational mixing is driven by meridional currents, which are inefficient at low rotational velocities. In contrast, in hydro models, shear mixing is not completely suppressed at low rotational velocities, as long as stars are subject to differential rotation. Even if rotational mixing is moderate for hydro models, it acts alongside overshoot in extending the MS evolution, which results in lower calibrated values of $\alpha_{\rm ov}$. Finally, models generally require higher $\alpha_{\rm ov}$ values to match the slow rotator TAMS line than the full sample TAMS line, as the former is located at lower temperatures. The only exception is the 12\,M$_\odot$ model, for which a weaker overshoot is required to fit the slow rotator TAMS, which is due to the fact that the two TAMS lines cross around $\log(L/\text{L}_\odot)\sim4.65$ (see Fig. \ref{TAMS_calib}).

\section{Impacts of stellar physics on the evolution}\label{stellar_tracks}
\subsection{Impacts of convective boundary mixing}
\begin{figure*}[h]
\centering
\centerline{\includegraphics[trim=.1cm 0.3cm .2cm .3cm, clip=true, width=2\columnwidth,angle=0]{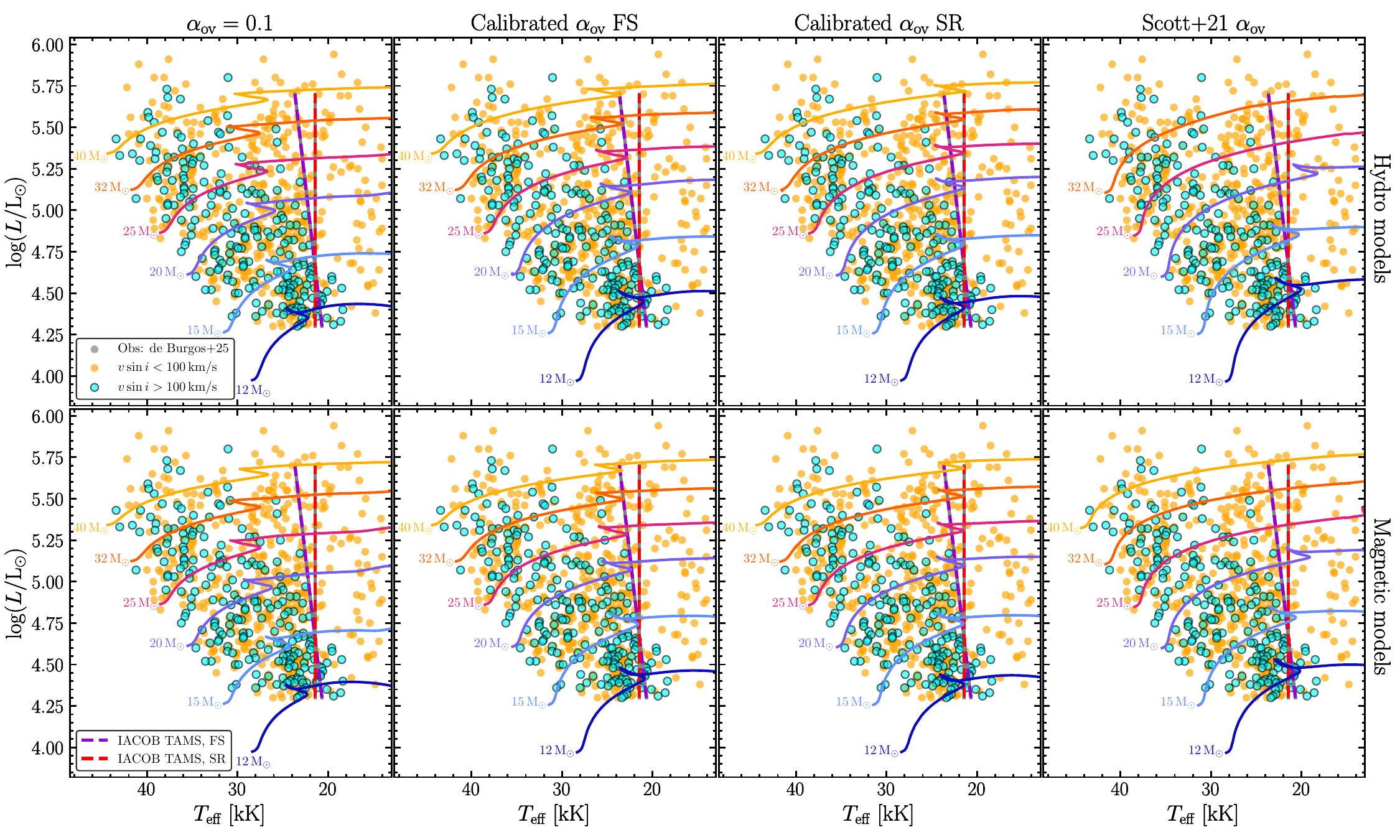}}
\caption{Hertzsprung-Russell diagram of the stellar models in the mass range $M_{\rm ini}=12-40$\,M$_\odot$ compared to observations from the IACOB sample \citep{deb25}. \textit{Upper panels:} Hydro models. \textit{Lower panels:} Magnetic models. \textit{Left to right:} $\alpha_{\rm ov}=0.1$, $\alpha_{\rm ov}$ calibrated with the full sample, $\alpha_{\rm ov}$ calibrated with the slow rotators, and $\alpha_{\rm ov}$ from \citet{sco21}. Stars with $\upsilon \sin i$ below and above 100\,km/s are respectively colored in orange and cyan. Purple-gray line: Full sample TAMS. Red-gray line: Slow rotator TAMS.}
\label{TAMS_calib}
\end{figure*}

The evolution in the HRD of stellar models computed under the choice of input physics and overshoot calibration described in Sect. \ref{ingredients} is shown in Fig. \ref{TAMS_calib} and compared to that of models with $\alpha_{\rm ov}=0.1$.\footnote{Note that these tracks differ from those of \citet{eks12} in their wind prescription and initial velocity.} Stars of the IACOB sample and the two TAMS lines (full sample and slow rotators) are overplotted. Additionally, we computed tracks with the $\alpha_{\rm ov}$ fitting formula by \citet{sco21}:
\begin{equation}
    \alpha_{\rm ov}=-0.00037867M_{\rm ini}^2+0.03885918M_{\rm ini}-0.01237484.
    \label{scott_eq}
\end{equation}
\citet{sco21} obtained Eq. \eqref{scott_eq} comparing the outputs of 1D stellar models with convective overshoot and entrainment laws. They provided scaled overshoot parameters as functions of initial mass, with different scaling relations depending on the power $m$ to which the convective velocity is raised. The $m=1$ case corresponds to a linear scaling between convective velocity and overshoot efficiency and is the author's preferred solution, for which they provide the fitting formula \eqref{scott_eq}. Here we test whether this theoretical prediction matches the observational constraints from \citet{deb25}. The corresponding $\alpha_{\rm ov}$ values are reported in Table \ref{table_overshoot}.
The following comments can be made regarding Fig. \ref{TAMS_calib}:
\begin{enumerate}
    \item Very similar evolutions are obtained with hydro and magnetic models. This is explained by the fact that at low velocity, rotational mixing only plays a minor role, and no significant differences are expected with various AM transport treatments. Additionally, for the full sample and slow rotator models, $\alpha_{\rm ov}$ is calibrated so that the models match the empirical TAMS. Therefore, the differences in rotational mixing efficiency are compensated for by the different overshoot efficiencies.
    \item By construction, models perfectly fit the empirical TAMS of the considered sample (full sample or slow rotators).
    \item Models with $\alpha_{\rm ov}=0.1$ are incompatible with both empirical TAMS lines. Their MS hook are situated at substantially too high temperatures, which suggests that CBM is underestimated in this case.
    \item Models computed with $\alpha_{\rm ov}$ efficiency from \citet{sco21} extend beyond the empirical TAMS lines. In particular, models with masses $M_{\rm ini}\sim 25-40$\,M$_\odot$ expand drastically, which is due to the fact that the fitting formula predicts very high overshoot efficiencies in this mass range ($\alpha_{\rm ov}\sim 0.72-0.94$).\footnote{Due to convergence issues, the $40$\,M$_\odot$ hydro model did not complete its MS and is not shown in Fig. \ref{TAMS_calib}.} As a result, their MS evolution is so prolonged that the MS hook is situated at effective temperatures $T_{\rm eff}<14$\,kK, which is the lower limit of the sample by \citet{deb25}. Thus, the overshoot efficiencies proposed by \citet{sco21} appear incompatible with the observational constraints from the IACOB sample for masses $M_{\rm ini}\gtrsim 20$\,M$_{\odot}$.
\end{enumerate}
\subsection{Impacts of rotation}\label{velocity_dependence}
Stars in the IACOB sample show diverse range of velocities. It has been shown by \citet{deb25} that the empirical TAMS lines differ when the full sample is restricted to slow rotators or fast rotators. The empirical TAMS line of the fast rotators subsample is well fit with the second-order polynomial:
\begin{equation}
    \log (L/\text{L}_\odot)=-0.001\,T_{\rm eff}^2+0.154\,T_{\rm eff}+1.71\,[\text{dex}].
    \label{TAMS_eq_FR}
\end{equation}Given that rotational mixing alters the MS expansion of the models, it is interesting to investigate whether the velocity dependence of the TAMS lines can be solely explained by single-star evolution.

\begin{figure}[h]
\centering
\centerline{\includegraphics[trim=.1cm 0.3cm .1cm .4cm, clip=true, width=1\columnwidth,angle=0]{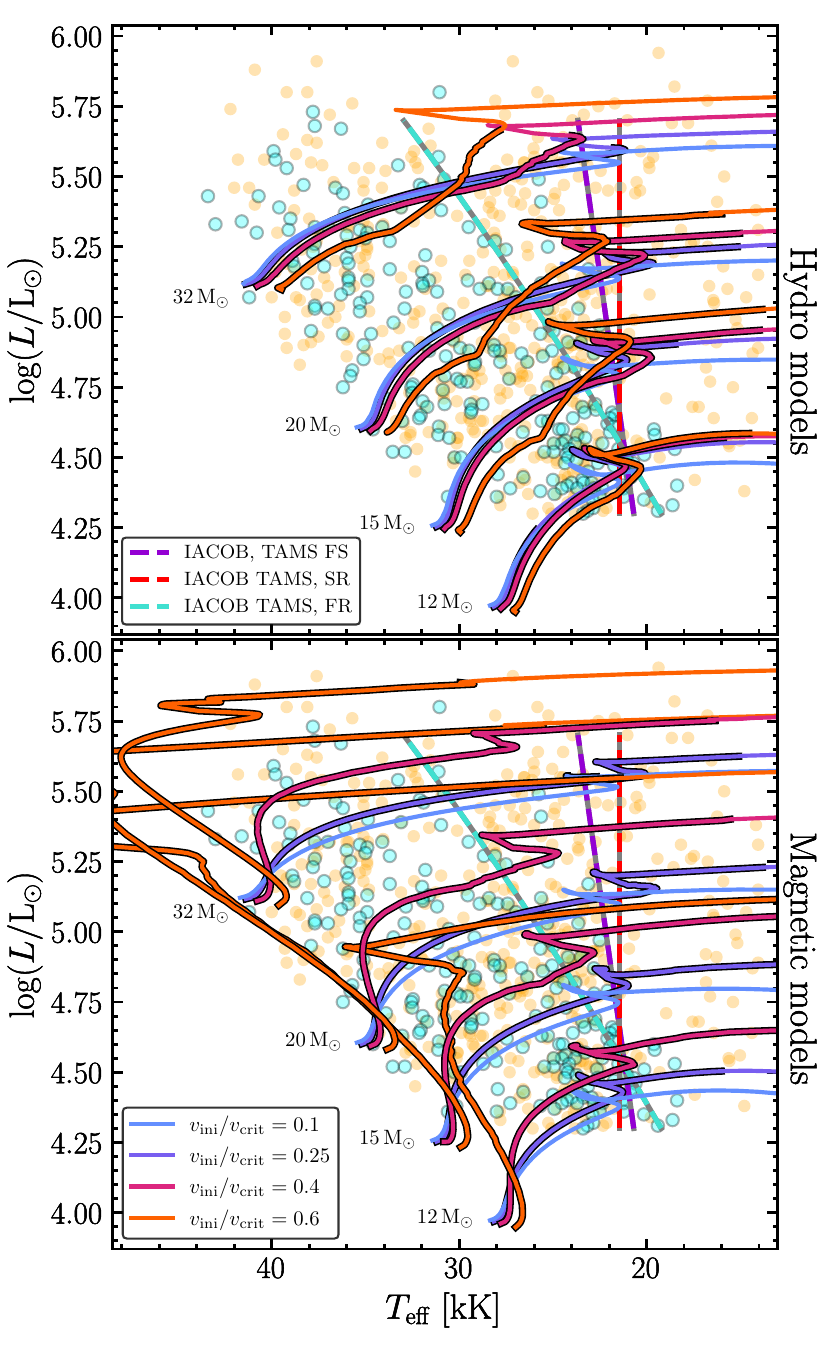}}
\caption{Hertzsprung-Russell diagram  of the stellar models in the mass range $M_{\rm ini}=12-32$\,M$_\odot$ with various initial rotational velocities compared to observations from \citet{deb25}. \textit{Top:} Hydro models. \textit{Bottom:} Magnetic models. Observed stars are color coded as in Fig. \ref{TAMS_calib}. Purple-gray line: Full sample TAMS. Red-gray line: Slow rotator TAMS. Cyan-gray line: Fast rotator TAMS. Black contours were added to the track when the star has \mbox{$\upsilon_{\rm eq}\braket{\sin i}>100\,$km/s}, with $\braket{\sin i}=\pi/4$.}
\label{HRD_vel}
\end{figure}

For this purpose, we computed grids of single-star models with initial velocities \mbox{$\upsilon_{\rm ini}/\upsilon_{\rm crit}=0.0,0.05,0.1,0.15,0.25,0.4,0.6$}\footnote{The critical velocity is computed as in \citet{eks12}, accounting for deformation caused by rotation.} for each AM transport treatment and overshoot calibration. This corresponds to angular velocities in the range $\Omega_{\rm ini}/\Omega_{\rm crit}=0-0.8$. In total, this corresponds to 6 (masses) $\times$ 7 (vel.) $\times$ 2 (calib.) $\times$ 2 (AM transport) = 168 models.\footnote{The tracks are available on \href{https://zenodo.org/records/22082876}{Zenodo}, \href{https://doi.org/10.26037/yareta:43xamo733jakpit6wwmyexuozi}{Yareta}, and on the \href{https://www.unige.ch/sciences/astro/evolution/en/database}{Geneva stellar group database}.} Figure \ref{HRD_vel} shows the evolution in the HRD of stellar tracks with initial velocities \mbox{$\upsilon_{\rm ini}/\upsilon_{\rm crit}=0.1,0.25,0.4,0.6$}, computed with the slow rotators $\alpha_{\rm ov}$ calibration. The tracks of the models with $\upsilon_{\rm ini}/\upsilon_{\rm crit} =0.0,0.05,0.15$ are not shown in this section but are used in the population synthesis (see Sect. \ref{population_comparison}). Here, we assume that the overshoot efficiency only depends on the initial mass, allowing the results of the calibration to be extrapolated to various initial rotational velocities. Stars of the IACOB sample and the three TAMS lines (full sample, slow rotators and fast rotators) are overplotted.

The following remarks can be made regarding Fig. \ref{HRD_vel}:
\begin{enumerate}
    \item The overall evolution, and in particular the TAMS location of the models is very sensitive to the initial velocity and AM transport treatment.
    \item Most models with $\upsilon_{\rm ini}/\upsilon_{\rm crit}\gtrsim0.4$ reach the critical velocity during the MS. In this case, mechanical mass loss makes the star evolve more vertically in the HRD, resulting in a noticeable angle in the tracks, altering the location of the MS hook.
    \item The minimum temperature reached by the models, which defines the theoretical TAMS, displays a non-monotonic trend with initial velocity. In general, the $\upsilon_{\rm ini}/\upsilon_{\rm crit}=0.25$ models expand more and end their MS at a lower $T_{\rm eff}$ than the $\upsilon_{\rm ini}/\upsilon_{\rm crit}=0.1$ models. This is due to the fact that rotational mixing increases the size of the convective core, which extends the MS lifetime and shifts the MS hook to cooler temperatures. However, at higher initial velocities ($\upsilon_{\rm ini}/\upsilon_{\rm crit}\gtrsim0.4$), the MS hook shifts to hotter temperatures, which is explained by the fact that mixing is efficient and models are more chemically homogeneous (see \citealt{far22} for a thorough discussion on how various processes drive the overall evolution of stellar models in the HRD), and the fact that most models reach the critical velocity during the MS.

    \item Magnetic models are more sensitive to the initial velocity. Models with $\upsilon_{\rm ini}/\upsilon_{\rm crit}=0.6$ experience quasi-chemically homogeneous evolution (CHE, see, e.g., \citealt{mae87,dem09,mar16}) and reach higher luminosities than corresponding models with only hydrodynamical instabilities. This is due to the fact that the strong AM transport keeps the star out of equilibrium with respect to meridional circulation, boosting its effect on rotational mixing. As a result the TAMS location of magnetic models with $\upsilon_{\rm ini}/\upsilon_{\rm crit}\gtrsim0.4$ is situated at hotter temperatures than that of hydro models.
    \item The fact that the black contours disappear during the evolution indicate that the models' surface velocity decreases. In particular, $32-40$\,M$_\odot$ hydro models become slow-rotators ($\upsilon_{\rm eq}\braket{\sin i}<100\,$km/s) before the end of the MS regardless of the initial velocity, due to the angular momentum removal by stellar winds and the moderate core-envelope coupling.\footnote{40\,M$_\odot$ and 32\,M$_\odot$ models behave similarly in this regard.} This is a crucial aspect to consider when comparing models to observations: the fact that the observed fast rotators and slow rotator TAMS lines do not coincide may simply be explained by the angular momentum removal by stellar winds, rather than an indication that the TAMS location is altered by rotation. However, with our choice of stellar winds prescription for OB stars \citep{bjo23}, this effect appears insufficient for altering the TAMS location, as all models born fast rotators only become slow rotators after their MS hook.\footnote{Except for the 32\,M$_\odot$, $\upsilon_{\rm ini}/\upsilon_{\rm crit}=0.6$ hydro model but the transition to slow rotator occurs just before the hook.} This result may be interpreted as an indication that angular momentum removal by stellar winds is underestimated with the \citet{bjo23} prescription. More efficient stellar winds may remove more angular momentum, making the stars transition from fast rotators to slow rotators earlier in their evolution at high masses ($M\gtrsim 25$\,M$_\odot$), which could explain the rotation dependence of the empirical TAMS lines. However, recent studies \citep[e.g.,][]{bjo21,bjo23,bra22,deb24,haw24,krt24} almost unanimously conclude that the \citet{vin01} prescription overestimates the strength of the winds, which undermines this hypothesis. Furthermore, it was shown by \citet{hol22} that \Genec hydro models computed with the \citet{vin01} prescription are incompatible with the rotational properties of the IACOB sample (see also Sect. \ref{rot_prop}).
\end{enumerate}
These results indicate that both types of models are unable to explain the velocity dependence of the empirical TAMS lines. For both types of AM transport treatment, the theoretical TAMS of the $\upsilon_{\rm ini}/\upsilon_{\rm crit}=0.4$ models, which correspond to the bulk of the fast-rotator population of the TAMS sample (see Fig. \ref{init_vel_distr}), are situated at cooler temperatures than the observed fast rotator TAMS. In the case of hydro models, the AM removal by stellar winds offers a promising explanation, but this effect appears insufficient to recover the empirical TAMS lines with recent, observation-calibrated mass loss rates. In the case of magnetic models, CHE makes the stars stay in the blue, shifting the TAMS location to hotter temperatures, but this only occurs for models with very high initial rotation. Indeed, the TAMS line of models initialized at $\upsilon_{\rm ini}/\upsilon_{\rm crit}=0.4$ is still situated at cooler temperatures than the empirical TAMS line, although this initial velocity corresponds to the bulk of the fast rotators observed population, and only $\sim15\,\%$ of stars ($\sim35\,\%$ of fast rotators) are found with higher velocities close to ZAMS.
\subsection{Rotational properties comparison}\label{rot_prop}
\begin{figure}[h]
\centering
\centerline{\includegraphics[trim=.3cm 0.3cm .3cm .3cm, clip=true, width=1\columnwidth,angle=0]{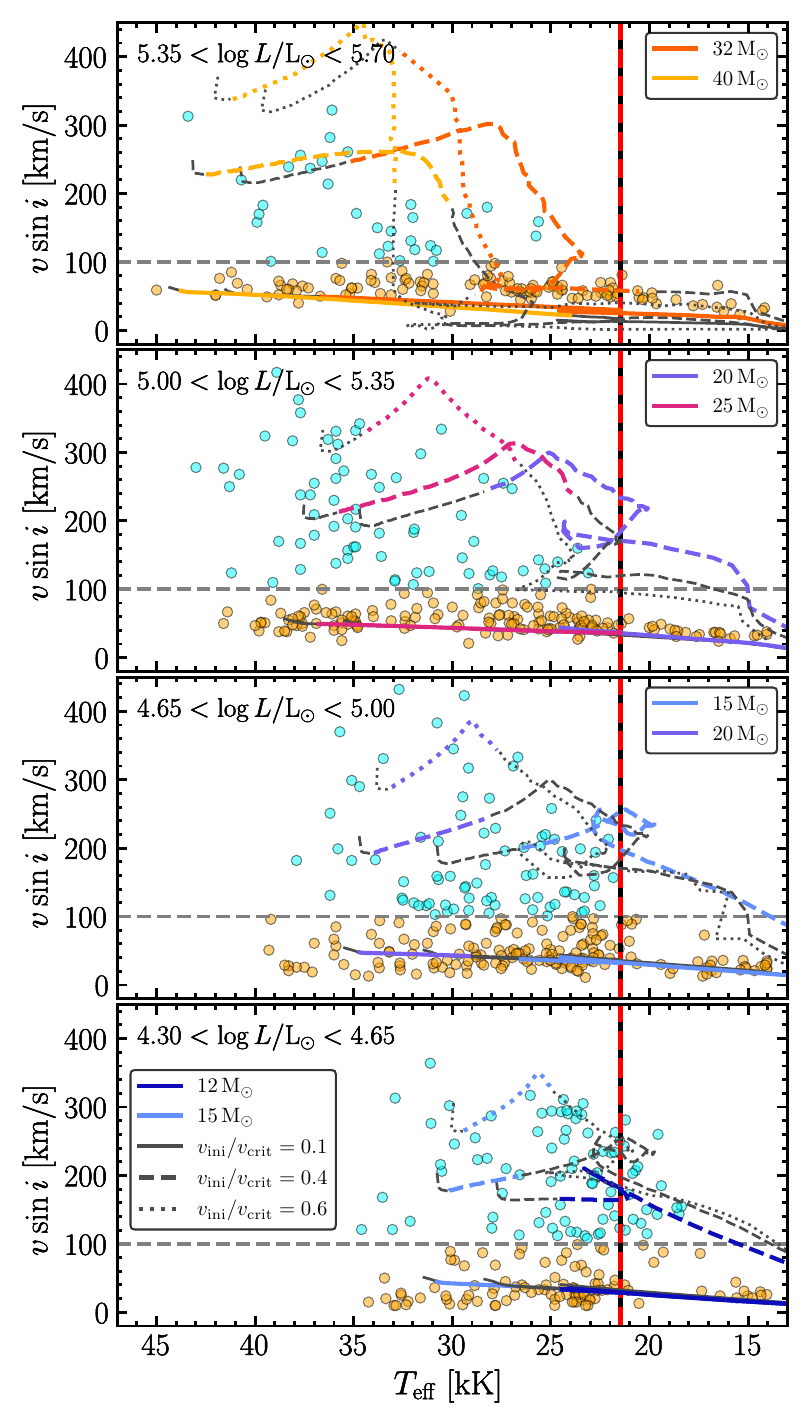}}
\caption{Temperature-velocity diagram of stellar models in the mass range $M_{\rm ini}=12-40$\,M$_\odot$ with initial rotational velocities of \mbox{$\upsilon_{\rm ini}/\upsilon_{\rm crit}=0.1,0.4,0.6$} compared to observations from \citet{deb25}. Observed stars are color coded as in Fig. \ref{TAMS_calib}. Vertical red and black line: Slow rotator TAMS. Dashed gray line: $\upsilon \sin i=100$\,km/s. Tracks outside the corresponding luminosity bin are plotted in dark gray.}
\label{velocity_teff_comp}
\end{figure}

In this section we compare the rotational properties of the models to those of the IACOB sample. As in \citet{deb25}, we show in Fig. \ref{velocity_teff_comp} the velocity-temperature diagram and compare observational data to stellar tracks with initial velocities $\upsilon_{\rm ini}/\upsilon_{\rm crit}=0.1,0.4,0.6$. This choice of initial velocities is motivated by the bimodality of the velocity distribution (see Fig. \ref{init_vel_distr}). The bulk of the slow-rotator and fast-rotator populations are represented by $\upsilon_{\rm ini}/\upsilon_{\rm crit}=0.1$ and $\upsilon_{\rm ini}/\upsilon_{\rm crit}=0.4$ models, respectively. The tracks shown in Fig. \ref{velocity_teff_comp} are those of hydro models with the slow rotator calibration. Similar results are obtained with magnetic models and the full sample calibration (see Fig. \ref{velocity_teff_syc}). Consequently, the rotational properties of the observed population do not allow us to constrain the AM transport efficiency.

Here, we test whether single-star evolution is able to reproduce the main rotational properties of the population. The aim is not to claim that the bimodality of the velocity distribution can be explained by single-star evolution (the growing consensus is that fast rotators are mainly due to binary interactions, e.g., \citealt{dem13,hol22,bri23}). This test merely assesses whether once formed, the evolution of slow rotators and fast rotators modeled as single stars is compatible with the observations. Even if the majority of fast rotators are indeed binary interaction products, it is possible that the companion does not affect much the evolution of the secondary once mass transfer stops, in which case its post-mass transfer evolution may be in broad agreement with single-star evolution.

As in \citet{deb25}, we split observational data in four different luminosity bins. 12 and 15\,M$_\odot$ models are compared to stars in the luminosity range $\log L/\text{L}_\odot = 4.30-4.65$, 15 and 20\,M$_\odot$ models for $\log L/\text{L}_\odot = 4.65-5.00$, 20 and 25\,M$_\odot$ models for $\log L/\text{L}_\odot = 5.00-5.35$, and 32 and 40\,M$_\odot$ models for $\log L/\text{L}_\odot = 5.35-5.70$. As in Fig. \ref{HRD_vel}, the velocities in the models were lowered by a factor of $\braket{\sin i}=\pi/4$. Throughout their evolution, models shift from one luminosity bin to the next. In each panel, tracks outside the corresponding bin are shown in dark gray.

The following comments can be made regarding Fig. \ref{velocity_teff_comp}:
\begin{enumerate}
    \item Models are in overall agreement with observations. The $\upsilon_{\rm ini}/\upsilon_{\rm crit}=0.1$ models reproduce both the trend and the average values of the slow rotator population.\footnote{In the two upper panels the models on average rotate slightly slower than the observed population; in the lower panel slightly faster. This is due to our choice of using one single initial velocity value for the overshoot calibration. Despite the mentioned deviations, this unique value proves to offer a satisfactory fit to the population of slow rotators at first order. In Sect. \ref{population_comparison} however, we use various initial velocities for generating synthetic populations.} The \mbox{$\upsilon_{\rm ini}/\upsilon_{\rm crit}=0.4$ and 0.6} models reproduce to first order the fast-rotator population.
    \item The predicted evolution of slow rotators and fast rotators significantly differ. Models with $\upsilon_{\rm ini}/\upsilon_{\rm crit}=0.1$ evolve almost at constant velocity (slight decrease), which is compatible with the observation. In contrast, $\upsilon_{\rm ini}/\upsilon_{\rm crit}=0.4$ and 0.6 models reach the critical velocity during the MS. When this happens, mechanical mass loss removes substantial amount of angular momentum, ensuring that the stars do not exceed the critical velocity. This effect reduces the velocity of the stars, which explains the observed hooks in the $T_{\rm eff}-\upsilon \sin i$ diagram. The predicted evolution of fast rotators seems consistent with the observations. The critical velocity consists in a physical limit, which manifests itself as a diagonal line in the $T_{\rm eff}-\upsilon \sin i$ diagram (decreasing velocity with decreasing temperature) and appears robust across masses. We note that almost no stars are observed beyond the region delimited by the tracks of models at the critical velocity. The few outliers are close enough to the tracks to be explained by inclination effects.
    \item Fast-rotating models fill the whole $T_{\rm eff}-\upsilon \sin i$ parameter space where stars are observed. In contrast, \citet{eks12} models experience significant spin-down, which was found to be incompatible with observations \citep{hol22}. We attribute this difference to the winds prescription, given that this ingredient is the main difference between the present models at $\upsilon_{\rm ini}/\upsilon_{\rm crit}=0.4$ and the \citet{eks12} models (notably the AM transport is the same; the treatment of overshoot also differs but it is not expected to significantly alter the rotational properties). The present models use the \citet{bjo23} prescription, whereas \citet{eks12} models use \citet{vin01}. When the \citet{bjo23} prescription is used, models are able to reproduce the observations even with a moderate core-envelope coupling. This indicates that the disagreement between the observations and the \citet{eks12} tracks observed by \citet{hol22} does not rule out moderate core-envelope coupling but rather indicates that the \citet{vin01} prescription overestimates mass loss in OB stars.
    \item \citet{deb25} observed a significant drop in the number of fast rotators beyond the TAMS line (i.e., for \mbox{$T_{\rm eff}<21.45$\,kK}, see also \citealt{vin10}). They noted that this aspect somewhat challenges single-star evolution. Indeed, the region forbidden by critical velocity extends to sensibly higher velocities than those of the observed population in this temperature range. Our results are compatible with this finding. This is particularly visible in the two central panels of Fig. \ref{velocity_teff_comp}. Our simulations indicate that fast rotators can stay at velocities between 100 and 200\,km/s after the TAMS, whereas no star is observed with these velocities in this temperature range. This conundrum, however, can be solved as follows: Even though the tracks cross the $T_{\rm eff}-\upsilon \sin i$ diagram in regions where no stars is observed, this transition occurs very rapidly such that it is in fact not expected statistically to observe many stars in this region when only single-star evolution is considered, as tested in \citet{deb25}. The actual problem of single-star evolution is that it in unsuccessful at explaining the population of slow rotators situated to the right of the TAMS line for the same reason, as is demonstrated in the following section.
\end{enumerate}
\section{Population comparison}\label{population_comparison}
In this section we extend the comparison between stellar models predictions and the observed population by generating synthetic populations from the four grids of stellar tracks (two AM transport treatments: hydro and magnetic models, two overshoot calibrations: slow rotators and full sample). For each set of stellar tracks, populations of 10.000 stars were generated using \Syclist \citep{geo14}, a population synthesis code designed specifically for \Genec tracks. To generate populations, we randomly draw for each star an age, initial mass, initial velocity and inclination from the distributions described in Sect. \ref{init_dist}. Stellar tracks are interpolated in mass and velocity. Whenever the age drawn is higher than the lifetime of the star, the star is not kept in the population. In this case, another star is generated with new random drawing of initial parameters.
\subsection{Initial distributions and main assumptions}\label{init_dist}
We assumed a constant star formation rate with a uniform initial mass function (IMF) in the mass range $M_{\rm ini}\sim 10-60$\,M$_\odot$. A uniform IMF may appear a questionable choice \citep[see, e.g.,][]{sal55,kro01}, but the observed population is rather uniform in mass. This is partly due to the \citet{mal22} bias, and the fact that binary interactions, which are not accounted for in this study, likely alter the evolutionary channels. As such, the observed population is not representative of a \citet{sal55} IMF. Moreover, our aim is to test whether single-star evolution can reproduce both the populations situated to the left and right sides of the TAMS line. Our result is sufficiently robust across the whole mass range that the IMF choice does not alter the outcome. The mass range is motivated by the luminosities covered in the study by \citet[][$\log L/\text{L}_\odot=4.3-5.7$]{deb25}. The luminosity range is completely covered with this choice of initial masses.

To obtain an initial velocity distribution, we used observational data from \citet{deb25} directly. We restricted the sample to surface gravities of $\log g_{\rm surf}>3.7$, as in \citet{hol22}, which corresponds to stars close to the ZAMS. We obtained a restricted sample of 152 stars, which we used to create an empirical cumulative distribution function (CDF). To obtain a $\upsilon/\upsilon_{\rm crit}$ distribution, observed velocities were raised by a factor $1/\braket{\sin i}=4/\pi$. The histogram of the empirical distribution is shown in Appendix \ref{AppB}. We note that the empirical distribution reproduces the bimodality of the distribution.

For inclinations, we assumed a probability distribution \mbox{$P(i)=\sin i$} as in \citet{gra88,geo14}. It naturally follows from the assumption of isotropic stellar orientation.

After interpolating the tracks, we accounted for gravity and limb darkening \citep{esp11,cla11} and corrected the temperatures and luminosities as in \citet{geo14}. Finally, to mimic the observational uncertainties, we added Gaussian noise to the compared quantities ($T_{\rm eff}, L$, and $\upsilon\sin i$). More details are given in Appendix \ref{AppC}.
\subsection{Distributions in the Hertzsprung-Russell diagram}\label{sect_HRD_syclist}
Figure \ref{HRD_syclist} shows the distributions of stars in the HRD of our synthetic populations. The following comments can be made based on the figure:
\begin{figure*}[h]
\centering
\centerline{\includegraphics[trim=.3cm 0.3cm .cm .3cm, clip=true, width=1.65\columnwidth,angle=0]{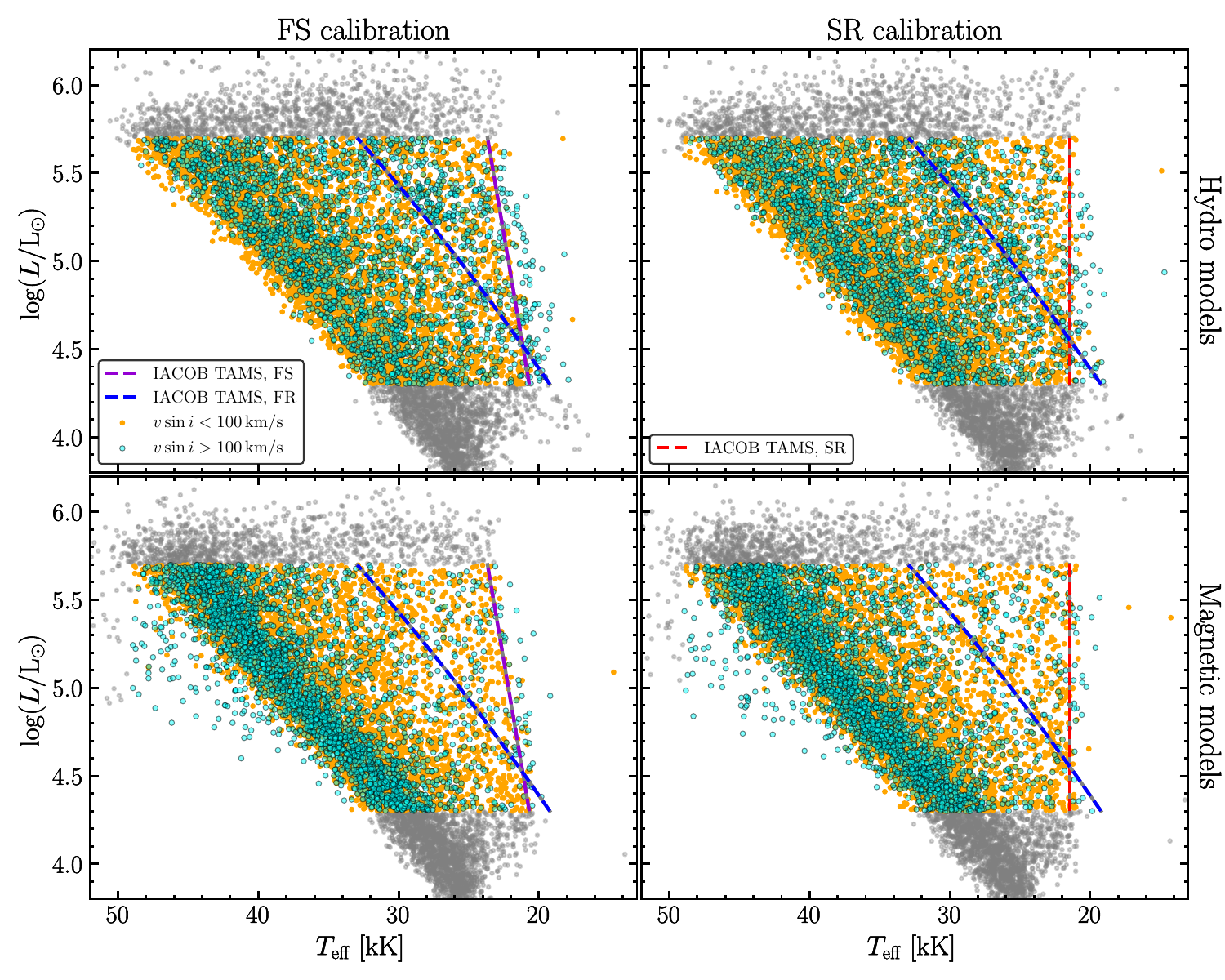}}
\caption{Hertzsprung-Russell diagram of the synthetic populations generated with \syclist. \textit{Upper panels:} Hydro models. \textit{Lower panels:} Magnetic models. \textit{Left panels:} Full sample $\alpha_{\rm ov}$ calibration. \textit{Right panels:} Slow rotators $\alpha_{\rm ov}$ calibration. Generated stars are color coded as in Fig. \ref{TAMS_calib}. Stars outside the temperature or luminosity range of the \citet{deb25} sample are shown in gray. Purple-gray line: Full sample TAMS. Red-gray line: Slow rotator TAMS. Blue-gray line: Fast rotator TAMS.}
\label{HRD_syclist}
\end{figure*}
\begin{enumerate}
    \item For each type of AM transport treatment, the population of slow rotators perfectly reproduces the empirical TAMS, being the full sample TAMS for the models computed with the ``full sample calibration'' or the slow rotator TAMS for those computed with the ``slow rotator calibration''. Indeed, the density of generated stars drastically drops at the empirical TAMS line. This validates our approach in matching the theoretical TAMS to the empirical TAMS.
    \item Several discrepancies can be observed between the synthetic and observed populations. The first of them is the population of stars near the ZAMS. Although models predict an overdensity of stars close to the ZAMS, it is not seen in the observed population. This discrepancy was already discussed in earlier studies. It is attributed mainly to two reasons: young stars are likely to be still embedded in their birth clouds, altering their observational properties \citep{sch21}; young stars probably keep accreting after the onset of hydrogen burning, altering the theoretical ZAMS \citep{hol20}.
    \item Independently of the AM transport and overshoot calibration, models are unsuccessful at recovering the velocity dependence of the empirical TAMS, as populations of fast rotators are observed beyond the empirical fast rotator TAMS. Fast rotators are even observed to the right of the slow rotator TAMS (respectively full sample TAMS for the models with the full sample calibration), where very few slow rotators are observed. These results are consistent with the tracks of fast-rotating stars (see Sect. \ref{velocity_dependence}). They highlight that single-star evolution, in particular the physics of rotation, appears insufficient in explaining the full observed distribution of stars in the HRD, or the necessity of recalibrating free parameters in rotating-models (AM transport and/or rotational mixing efficiency).
    \item Magnetic models predict populations of fast rotators to the left of the ZAMS, which is due to the fact that $\upsilon_{\rm ini}/\upsilon_{\rm crit}\gtrsim 0.6$ models are predicted to experience quasi-CHE. These stars typically have higher surface gravities than the upper limit of the \citet{deb25} sample ($\log g_{\rm surf}=4.2$), and would be observationally classified as hot subdwarfs. Thus, their absence in the \citet{deb25} sample does not rule out the rotational mixing efficiency of magnetic models. Given the opposite predictions of hydro and magnetic models  concerning this population (the population is not predicted by hydro models), observed populations of hot subdwarfs could be used in the future to constrain the rotational mixing efficiency of these models. This aspect is discussed in more details in Appendix \ref{AppD}.
    \item The most striking discrepancy between the observed and synthetic distributions is the population of stars to the right of the TAMS line. While in the observed population around 15\,\% of stars are found to the right of the TAMS line, in the synthetic population the fraction is $\sim 2$\,\%. Some fast rotators are observed to the right of the TAMS line in the synthetic populations, but we interpret this population as being made of stars still in their MS and that have an extended MS width due to the effects of rotational mixing. In contrast, in the observed population, the large majority of stars to the right of the TAMS line are slow rotators, which are very scarce in the synthetic populations. The paucity of stars to the right of the TAMS line is explained by the fact that the models rapidly cross the Hertzsprung gap after the end of core hydrogen burning phase. They only spend between 5 and 20\,kyr in this region before reaching temperatures cooler than 14\,kK, which corresponds to the lower limit of the \citet{deb25} sample. This duration typically corresponds to only 0.1\,\% of the star's lifetime, which explains the low percentage of stars predicted in this region. Most stars to the right of the TAMS line in the synthetic population are in fact fast rotators with a prolonged MS or slow rotators shifted to cooler temperatures by the synthetic noise. This result implies that single-star evolution is unsuccessful at explaining the population of stars to the right of the TAMS line in the IACOB sample. Earlier studies already suggested that this population could only be explained by binary interaction channels, notably mergers \citep[see, e.g.,][]{men24}. Our findings align with these results and strengthen them, demonstrating that the problem persists with various AM transport efficiencies, angle of view and measurement errors considerations.  We further discuss their robustness in Sect. \ref{discussion}.
\end{enumerate}

\subsection{Temperature-velocity distributions}
\begin{figure*}[h]
\centering
\centerline{\includegraphics[trim=.3cm 0.3cm .cm .3cm, clip=true, width=1.5\columnwidth,angle=0]{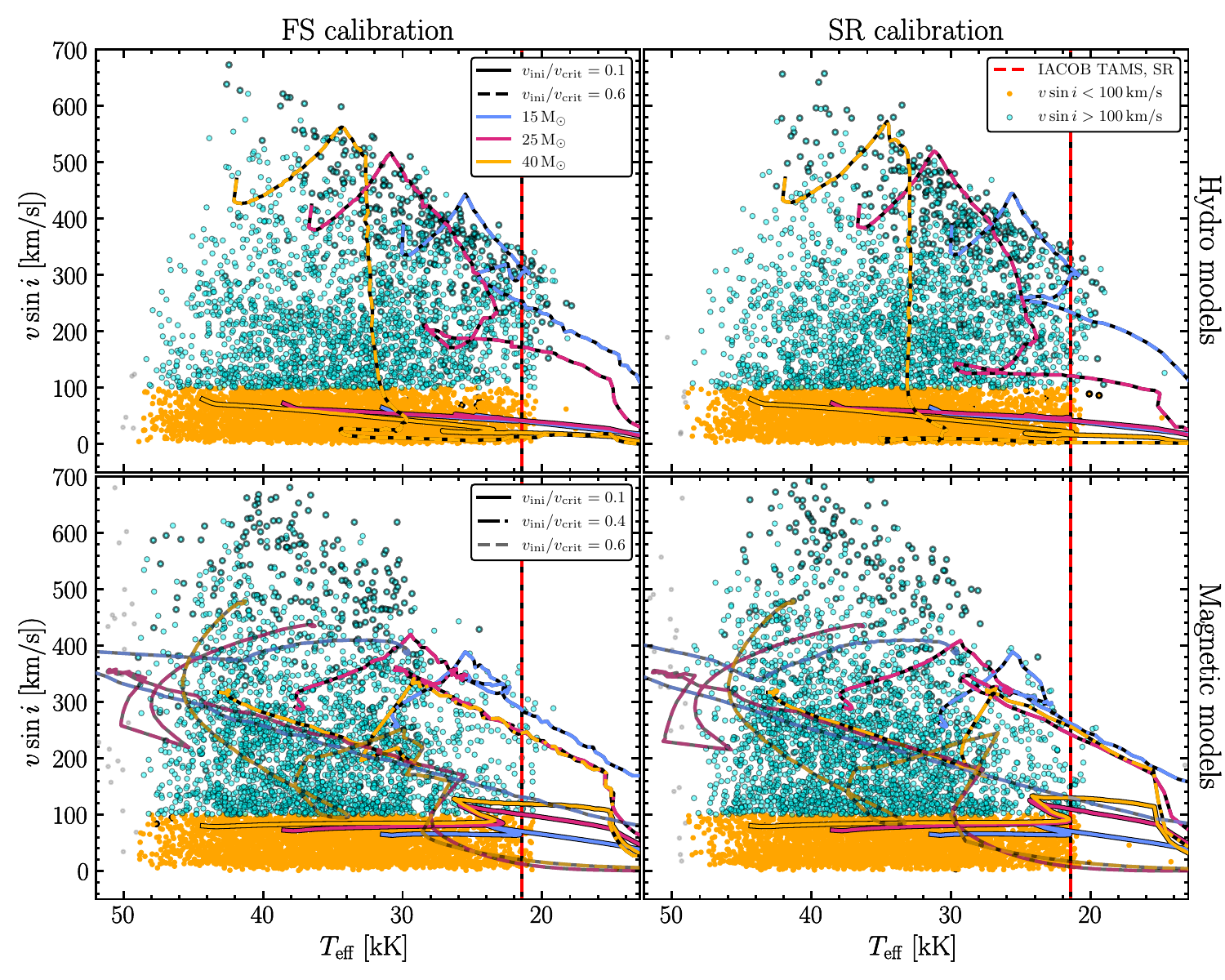}}
\caption{Temperature $-$ projected velocity diagram of synthetic populations generated with \syclist. \textit{Upper panels:} Hydro models. \textit{Lower panels:} Magnetic models. \textit{Left panels:} Full sample $\alpha_{\rm ov}$ calibration. \textit{Right panels:} Slow rotator $\alpha_{\rm ov}$ calibration. Generated stars are color coded as in Fig. \ref{TAMS_calib}. Thicker contours are used when the velocity was reduced to the critical velocity. Stars outside the temperature range of the \citet{deb25} sample are shown in gray. Red-black line: Slow rotator empirical TAMS. Stellar tracks are overplotted.}
\label{velocity_teff_syc}
\end{figure*}

Figure \ref{velocity_teff_syc} shows the $T_{\rm eff}-\upsilon\sin i$ diagram of the synthetic populations with stellar tracks overplotted. Due to extrapolations when the drawn initial velocity was higher than 0.6,\footnote{Above this value, models become computationally challenging They quickly reach the critical velocity and/or experience quasi-CHE. We recall that velocities $\upsilon_{\rm ini}/\upsilon_{\rm crit}=0.6$ correspond to angular velocities $\Omega_{\rm ini}/\Omega_{\rm crit}=0.8$.} some stars (about 4\,\% of the total population) were predicted to rotate above the critical velocity. In this case we corrected their equatorial velocity to ensure that the rotation is not super-critical. These stars are indicated with a thicker contour in Fig. \ref{velocity_teff_syc}.

Similar results are obtained for the two different AM transport treatments and overshoot calibrations. The main difference between hydro and magnetic models is that the latter predict more fast rotators at higher temperatures because they experience more mixing. In all of the synthetic populations, we observe a clear trend in the fast-rotator population: The maximum velocity decreases with decreasing temperatures, which can be attributed to the limit imposed by the critical velocity. Indeed, the trend well matches the tracks of models evolving at the critical velocity.\footnote{The magnetic models initialized at $\upsilon_{\rm ini}/\upsilon_{\rm crit}=0.6$ experience quasi-CHE and as a result do not reach the critical velocity during their MS. For this reason $\upsilon_{\rm ini}/\upsilon_{\rm crit}=0.6$ models are displayed as semi-transparent lines and models with $\upsilon_{\rm ini}/\upsilon_{\rm crit}=0.4$, which reach the critical velocity during their MS, are also shown in this case.} These results offer a theoretical explanation for the shape of the observed fast-rotator population in the $T_{\rm eff}-\upsilon\sin i$ diagram. Even though the majority of the fast rotators likely get their fast rotations from binary interactions, this physical limit is expected to also apply to binary systems.

Regarding the drop in the number of fast rotators beyond the empirical TAMS, we note here that single-star models do predict a significant drop of the fast-rotator population in the $T_{\rm eff}-\upsilon\sin i$ diagram, but it does not exactly match the location of the empirical TAMS. We attribute this result to the fact that fast-rotating models have an extended MS lifetime, which shifts their TAMS location to cooler temperatures. Nonetheless, the observed drop in their population can be attributed to their fast post-MS evolution, as proposed in Sect. \ref{rot_prop}. Consistently with this result, almost no slow rotators are predicted beyond the TAMS line. Ultimately, the main mismatches between the observed and synthetic populations remain the same: The physics of rotation is unsuccessful at explaining the empirical TAMS rotation dependence (the TAMS location of fast rotators is predicted at cooler temperatures) and single-star physics is unsuccessful at explaining the population of slow rotator BSGs beyond the TAMS location.
\section{Discussion}\label{discussion}
\subsection{Convective boundary criterion}\label{CBC_disc}
In all simulations the post-MS expansion is very fast. All the computed models only spend between 5 and 20\,kyr crossing the region to the right of the TAMS line up to temperatures cooler than 14\,kK, which is the lower limit of the \citet{deb25} sample. Regarding the possibility that this scenario changes using the Schwarzschild or Ledoux criterion for convective boundary, \citet{sib23} showed that the crossing of the Hertzsprung gap is systematically faster when the Ledoux criterion is used. We therefore expect the discrepancy between the observed and synthetic populations to the right of the TAMS line to be even more pronounced if the models are computed with the Ledoux criterion instead of the Schwarzchild criterion.
\subsection{Definitions of the TAMS lines}\label{definition_disc}
One key question is whether the discrepancy between the observed and synthetic populations is merely a consequence of the different ways empirical and theoretical TAMS are defined. In this work, we defined the theoretical TAMS using the minimum $T_{\rm eff}$ reached by the models during the MS. \citet{deb25} instead defined the empirical TAMS as the temperature at which the CDF of the observed population in four different luminosity bins drops below 15\,\%. They performed polynomial interpolations out of the four points to obtain empirical TAMS lines. The value of 15\,\% is motivated by the fact that in most luminosity bins it coincides with an actual drop of density in the HRD. One may wonder whether it is possible to calibrate the overshoot efficiency as to directly match synthetic and observed populations, without considering the empirical TAMS.

\begin{figure}[h]
\centering
\centerline{\includegraphics[trim=0.8cm 0.1cm 1.8cm 1.6cm, clip=true, width=1\columnwidth,angle=0]{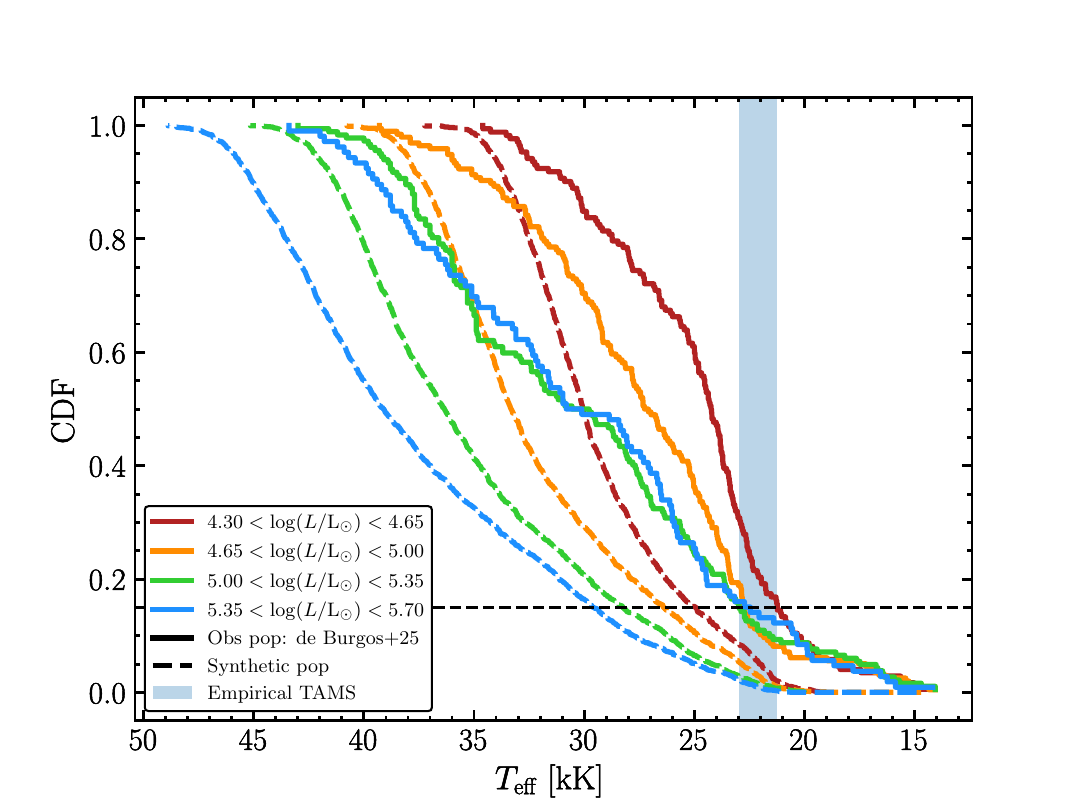}}
\caption{Observational and theoretical cumulative distributions in four luminosity bins. The blue shaded area corresponds to the empirical TAMS location, i.e. to the $T_{\rm eff}$ where the CDF of the different luminosity bins drop below 15\,\% (dashed black line).}
\label{CDF_comparison}
\end{figure}

To address this question, we compare in Fig. \ref{CDF_comparison} the CDFs of our preferred synthetic population, hydro models with the slow rotator calibration, to those of the observed population.
The discrepancy between the observed and synthetic populations is clearly illustrated. The blue shaded region indicates the location of the empirical TAMS, where the CDFs of the observed population drop below 15\,\%. In contrast, the CDFs of the models drop below 15\,\% at hotter temperatures, especially in the highest luminosity bins. We further notice that the CDFs of the models are already close to zero in the region of the empirical TAMS, which is consistent with the observed distribution in the HRD. In other words, by construction the theoretical TAMS line of the models matches the empirical TAMS line but not the observed CDFs. Defining the ``empirical TAMS line of the models'' with the $T_{\rm eff}$ where the CDFs drop below 15\,\% results in a discrepancy with the actual empirical TAMS line.

We conclude that an alternative calibration of the overshoot efficiency does not solve the problem. First, the shapes of the CDFs are quite different. Even with an alternative calibration, we expect the models to be unsuccessful at matching the empirical CDFs, whose slopes flatten below $\sim$\,21\,kK, due to the drop of density of the stars in the HRD. There is no reason for the density of the generated distributions to show similar drops. Second, even if a better match between the observed and simulated CDFs is possible, we stress that the large majority ($\gtrsim99.9$\,\%) of the simulated stars would be in their MS, as demonstrated by the short timescale over which they expand to $T_{\rm eff}<14$\,kK after core hydrogen exhaustion (about 0.1\,\% of their MS lifetime). This would lead to the following paradox: CDFs and empirical TAMS match, but almost all the stars to the right of the empirical TAMS are in fact still in their MS. We therefore claim that the mismatch between the observed and synthetic populations cannot be solely attributed to inconsistencies in the way the TAMS lines are defined but rather indicates that the observed population cannot be fully explained by single-star evolution.
\subsection{RSG and post-RSG phase evolution}\label{RSG_disc}
In this study the evolution of the models were stopped after reaching $T_{\rm eff}<14$\,kK. After the end of the MS, stars rapidly cross the Hertzsprung gap and become RSGs. It is possible that some stars return to the blue after the RSG phase, which was not accounted for in our models. This question is heavily debated \citep[see, e.g.,][]{bea21,mas23,van25}. The transition to the blue after the RSG phase is very sensitive to the strength of the winds experienced during this phase, which are very uncertain. As noted by \citet{deb25}, if a significant fraction of stars in their sample are indeed post-RSG BSGs, this would bias the location of the empirical TAMS line. Our models are unsuccessful at reproducing the observed population to the right of the TAMS line when the post-RSG phase is not accounted for. In this section we investigate whether single-star evolution can reproduce it through this channel.

\begin{figure}[h]
\centering
\centerline{\includegraphics[trim=0.3cm 0.3cm 0.cm 0.4cm, clip=true, width=1\columnwidth,angle=0]{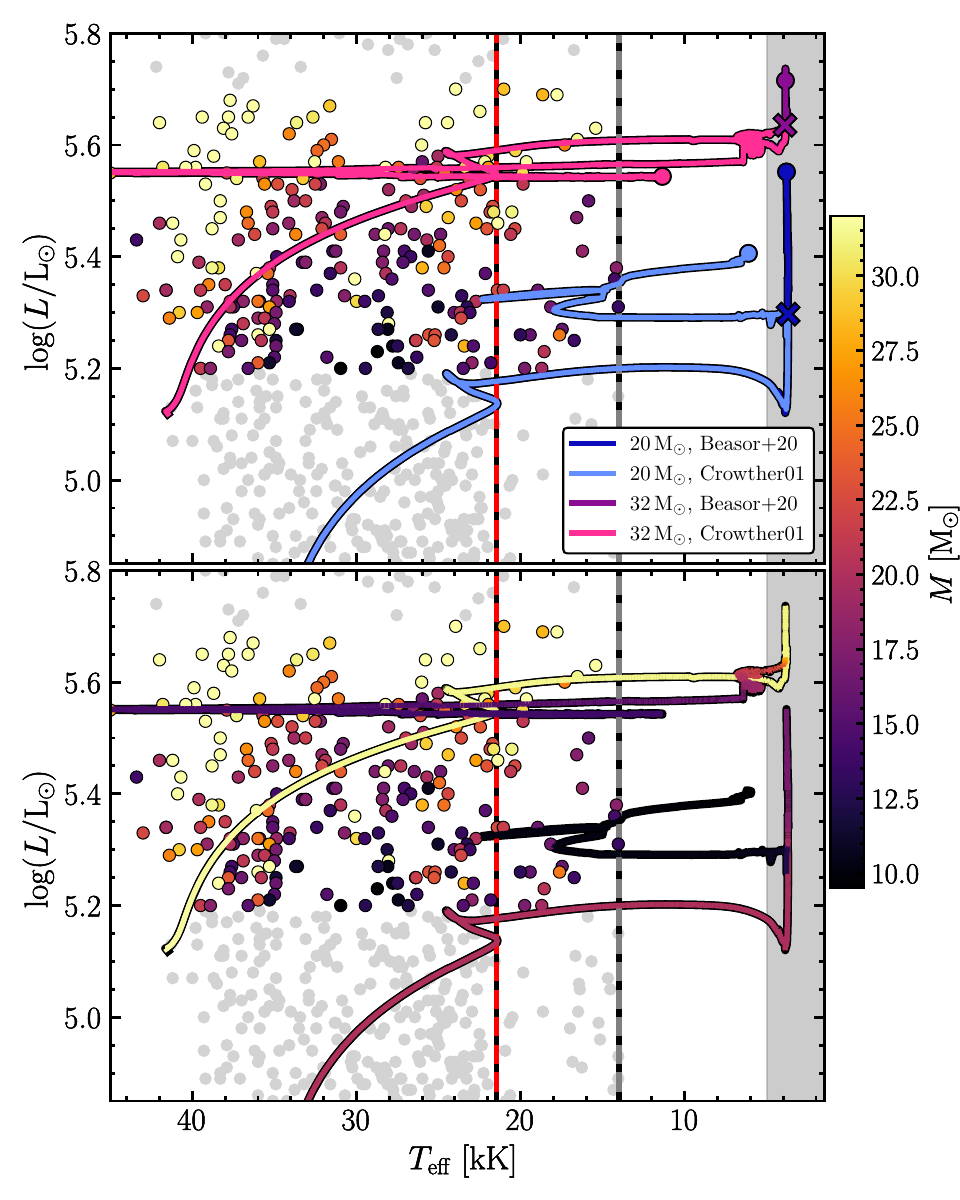}}
\caption{Evolution in the HRD of stellar models of masses \mbox{$M_{\rm ini}=20$ and $32$\,M$_\odot$} with different RSG wind prescriptions compared to observations from \citep{deb25}. \textit{Top:} Observations are color coded by their mass. The moment of the evolution when models computed with different wind prescriptions start to diverge is indicated with a cross, the last computed models with a dot. \textit{Bottom:} Models and observations are color coded by their mass. Red-black line: Slow rotator TAMS. Gray-black line: Temperature limit of the sample. Stars outside the luminosity bin \mbox{$\log L/\text{L}_\odot = 5.20-5.70$} are plotted in gray. Shaded gray region: \mbox{$\log(T_{\rm eff}$ [K]$)<3.7$}, where RSG winds apply.}
\label{HRD_RSG}
\end{figure}

In the \citet{eks12} grid, models above 20\,M$_\odot$ return to the blue after the RSG phase. Given the upper luminosity of the sample, we pursue the evolution of the $\upsilon_{\rm ini}/\upsilon_{\rm crit}=0.1$ models of 20 and 32 M$_{\odot}$ beyond the TAMS to obtain RSG models and study their behavior. To investigate the effect of the strength of RSG stellar winds, we compute models with the default \Genec RSG winds treatment: the \citet{cro01,eks12} prescription, and also with the recent \citet{bea20} prescription. The former typically predicts stronger winds. The evolution in the HRD of the four models is shown in Fig. \ref{HRD_RSG}. Observational data are overplotted and color coded by mass.\footnote{For observational data, we use spectroscopic masses derived by \citet{deb25} from radii (obtained following \citealt{her92}) and surface gravities (corrected for the centrifugal force following \citealt{rep04}).} In the lower panel of Fig. \ref{HRD_RSG}, the tracks are also color coded by mass, using the same scale as for the observations.

We note that when the \citet{bea20} prescription is used, models lose less mass. As a result, they stay RSGs until the end of their evolution (here the models were stopped at the end of carbon burning phase). As a result, they predict that the sample is not polluted by post-RSG BSGs. In contrast, when the \citet{cro01,eks12} prescription is used, the stars lose more mass, making them return to the blue. Both the 20 and 32\,M$_\odot$ models enter the region covered by the \citet{deb25} sample and stay there during respectively 417 and 224\,kyr. However, when this occurs stars have already lost significant amount of mass. When they reach $T_{\rm eff}>14$\,kK, the 20\,M$_\odot$ model only has kept 10.3\,M$_\odot$, while the 32\,M$_\odot$ model 15.7\,M$_\odot$.

If post-RSG evolution is the main channel for explaining the population of BSGs to the right of the TAMS line, according to the models we would expect to observe a significant mass discrepancy between the populations on the two sides of the TAMS line, given that the population situated on the left is dominated by MS stars. However, this is not observed using all the stars within the luminosity bin \mbox{$\log L/\text{L}_\odot = 5.20-5.70$}, where models return to the blue after the RSG phase. The median and average value of the spectroscopic masses to the left of the TAMS line are \mbox{$\widetilde M_{\rm sp}(T_{\rm eff}>14\mbox{\,kK})=21.1$\,M$_\odot$} and \mbox{$\overline{M}_{\rm sp}(T_{\rm eff}>14\mbox{\,kK})=23.6$\,M$_\odot$}. To the right of the TAMS line, they are \mbox{$\widetilde M_{\rm sp}(T_{\rm eff}<14\mbox{\,kK})=20.9$\,M$_\odot$} and \mbox{$\overline{M}_{\rm sp}(T_{\rm eff}<14\mbox{\,kK})=22.5$\,M$_\odot$}. In contrast, post-RSG models predict at least a $\sim50$\,\% difference. This discrepancy is also seen in Fig. \ref{HRD_RSG}, as post-RSG models appear less massive than observed stars in the same region of the HRD. The potential mass discrepancy between evolutionary and spectroscopic mass \citep[see, e.g.,][]{her92,mah15,mar18,bes20} does not solve the discrepancy between post-RSG models and observations, as it equally affects stars on each side of the TAMS line. This argument does not rule out the possibility that some stars in the IACOB sample are indeed post-RSG BSGs. Yet, it demonstrates that this channel is unsuccessful at explaining the population of BSGs to the right of the TAMS line.
\subsection{Main caveats}\label{caveats}
The main limitations of our models are the following:
\begin{enumerate}
    \item The grids rely on the assumption that the overshoot efficiency only depends on the initial mass. In reality, it is possible that overshoot interplays with the physics of rotation in a more complicated way such that the assumption that its efficiency is rotation independent is not valid. This would certainly alter the results of the fast-rotating models. Nonetheless, this assumption is commonly applied in widely used grids \citep[e.g.,][]{bro11a,eks12,cho16}.\footnote{These grids further assume a mass-independent overshoot efficiency.}
    \item We did not investigate overshoot efficiencies higher than those predicted by \citet{sco21} in this mass range (up to $\alpha_{\rm ov}=0.94$). In the investigated range of efficiencies, we find that the MS width monotonously increases with $\alpha_{\rm ov}$, which guarantees the uniqueness of the calibration. Yet, it is possible that above this threshold models react differently to increasing overshoot efficiencies, in which case other calibrations compatible with the observations may exist.
    \item Massive stars are estimated to form approximately 70\,\% of the time in binary or multiple systems \citep{san12,moe17}. It is expected that a nonnegligible fraction of the \citet{deb25} sample are binary products, even among the slow rotator population (see, e.g., \citealt{mar25,mar26,sim26b} for the case of O-type stars). The evolutionary tracks are drastically altered by binary interactions \citep[e.g.,][]{eld17,fra23}, and higher-order interactions further complicate the situation \citep[e.g.,][]{too16,pre24,sci25,bru25}. The overshoot calibrations proposed in this work are performed under the simplifying assumption that stars evolve as single stars. Accounting for binary interactions may alter the overshoot calibration. Nevertheless, we consider that slow rotators and single-star models are still the best available options for calibrating free parameters of stellar evolution.
    \item We stress that the results of the calibration of magnetic models should be used with caution, if the physics of rotational mixing differs. Magnetic models with a different physics, for instance, using the so-called $f_{c}$ -- $f_\mu$ treatment \citep{heg00a}, as is the case in \mesa models \citep{pax13} are expected to have similar angular velocity profiles (close to solid-body rotation due to the efficient AM transport), but not necessarily the same abundance profiles, as chemical mixing is treated differently. As the theoretical TAMS is affected by the combined effect of overshoot and rotational mixing, it is not guaranteed that the calibration of our magnetic models is valid for models computed with the $f_{c}$ -- $f_\mu$ treatment. However, given the limited differences between the results of hydro and magnetic models, we expect the calibration to be usable to first order.
    \item Our calibration indicates that the overshoot efficiency is mass dependent.\footnote{If parametrized with respect to the pressure scale height $H_P$.} It may be necessary, when computing large grids of single or binary stars, to adopt a single value for the overshoot efficiency. In this case, we suggest computing an average value out of the results of Table \ref{table_overshoot}, which is around $\alpha_{\rm ov}=0.3-0.35$, depending on the considered sample and AM transport treatment. It is worth noting that these values are in agreement with that used in the BONN models (0.335; \citealt{bro11a}). But as pointed out by \citet{deb25}, it leads to a mismatch between the empirical and theoretical TAMS for initial masses $M_{\rm ini}>25$\,M$_\odot$, which we now interpret as the sign that lower overshoot efficiencies are required in this mass range.
\end{enumerate}
\section{Conclusion}\label{conclusion}
In this study we have calibrated the overshoot efficiency to reproduce the empirical TAMS locations obtained by \citet{deb25} from the IACOB sample of O-type stars and B giants and supergiants using either the full sample or slow rotators only. We computed \Genec models with two different AM transport treatments (purely hydrodynamic or magnetohydrodynamic models). We performed four independent calibrations to reproduce with each AM transport treatment the position of the two empirical TAMS. The calibrations were performed with $\upsilon_{\rm ini}/\upsilon_{\rm crit}=0.1$ models (which corresponds to the bulk of the slow rotator population) in the mass range $M_{\rm ini}=12-40$\,M$_\odot$. The calibrated models accurately reproduce the observed population. Their theoretical TAMS lines perfectly align with the empirical TAMS lines. We find that matching the empirical TAMS lines requires mass-dependent overshoot efficiencies with values in the range $\alpha_{\rm ov}=0.18-0.45$, depending on the considered sample (slow rotators or full sample) and AM transport. These overshoot efficiencies are in good agreement with those obtained by \citet{lec26} using \mesa models and a Bayesian inference approach. The calibrated $\alpha_{\rm ov}$ values display non-monotonic trends with mass, with a peak around $M_{\rm ini}\approx15-20$\,M$_{\odot}$. This result appears to be in contradiction with \citet{sco21}, who found that the overshoot efficiency monotonously increases with initial mass.

Assuming that the overshoot efficiency is only mass dependent, we computed grids of \Genec stellar tracks with the two AM transport treatments and the different overshoot calibrations. We studied the effects of rotation by computing models with initial velocities $\upsilon_{\rm ini}/\upsilon_{\rm crit}=0.0,0.05,0.1,0.15,0.25,0.4$, and 0.6. We compared the tracks of rotating models and found that the effects of rotation are unsuccessful at reproducing the velocity dependence of the empirical TAMS by \citet{deb25}. These authors observed that the distribution of fast rotators in the HRD differs from the rest of the sample. Our investigations suggest that single-star evolution alone (including the effects of rotation) cannot explain the distribution of fast rotators unless free parameters (rotational mixing or AM transport efficiency) are recalibrated. In our view, this result rather reinforces the growing consensus that the fast-rotator population is dominated by binary interaction products \citep[e.g.,][]{dem13,hol22,deb25} and cannot be solely explained by single-star evolution.

We also compared the rotational properties of the stellar tracks to those of the observed population. We find that overall the models reproduce the observed rotational properties. The slow rotators seem to be compatible with an almost constant velocity evolution, as was observed by \citet{hol22}. The $\upsilon_{\rm ini}/\upsilon_{\rm crit}=0.1$ models reproduce this trend well. Fast-rotating models are able to fill the whole region of the $T_{\rm eff}-\upsilon\sin i$ parameter space where stars are observed. Their velocity typically increase with a decreasing $T_{\rm eff}$ until they reach the critical velocity, which makes their velocity decrease. The critical velocity appears to delimit a frontier in this diagram below which stars cannot evolve. This theoretical result is supported by the observations, as almost no star is observed beyond the region delimited by the path of the stars at critical velocity. This offers a theoretically supported explanation for the trend observed in the fast-rotator populations (i.e., maximum velocities decrease with $T_{\rm eff}$). In this regard, our models differ from those of \citet{eks12}, whose velocity evolution was found to be inconsistent with the rotational properties of the IACOB sample of O-type stars (\citealt{hol22}; see also \citealt{deb24} for the extension to the B supergiant domain). We now attribute this discrepancy to the OB winds. \citet{eks12} used the \citet{vin01} prescription, whereas in this study we used the \citet{bjo23} prescription, which predicts weaker winds.

We generated synthetic populations from the stellar tracks using \Syclist \citep{geo14} and compared them with the observed population. The synthetic populations validated the performed calibrations, as the density of stars in the HRD drastically dropped at the location of the empirical TAMS. However, we observed several discrepancies between the observed and synthetic populations. The synthetic populations contain large fractions of stars near the ZAMS that are not present in the observed population. This discrepancy has also been reported in earlier studies \citep[e.g.,][]{hol20,sch21}. As expected from the stellar tracks, fast rotators are predicted in regions where they are not observed. The most striking discrepancy between the  observed and synthetic populations is the fraction of stars to the right of the TAMS line. While 15\,\% of the observed stars are found in this region, the models predict a fraction of only about 2\,\%. This is explained by the fast expansion of the models after the end of the MS. Single-star models are unsuccessful at reproducing the population to the right of the TAMS line, even accounting for synthetic noise to mimic measurement uncertainties in the observed population; with alternative AM transport efficiencies or convective boundary criteria; when attempting to match the CDFs of the populations instead of their TAMS lines or when RSG and post-RSG evolution is considered. These results imply that the majority of stars to the right of the TAMS line in the \citet{deb25} sample cannot be explained by single-star evolution and are mostly binary-interaction products, notably mergers (see, e.g., \citealt{men24,sim26b}, de Burgos et al. in prep).

The calibration of the overshoot efficiency represents a crucial step toward next-generation stellar tracks. Given the results of this study, it appears that the widely adopted assumption of mass-independent overshoot efficiency is not supported by observations in the luminosity domain $\log L/\text{L}_\odot=4.30-5.70$.
\section*{Data availability}
The tracks are available on \href{https://zenodo.org/records/22082876}{Zenodo}, \href{https://doi.org/10.26037/yareta:43xamo733jakpit6wwmyexuozi}{Yareta}, and on the \href{https://www.unige.ch/sciences/astro/evolution/en/database}{Geneva stellar group database}.
\begin{acknowledgements}
We sincerely thank the anonymous referee for their constructive feedback and valuable suggestions, which enabled us to substantially improve the manuscript. L.S., S.R. and S.E. acknowledge support from the SNF project No 212143. S.S-D. acknowledges support from the State Research Agency (AEI) of the Spanish Ministry of Science and Innovation (MICIN) and the European Regional Development Fund, FEDER under grant PID2024-159329NB C21. The project leading to this application has received funding from European Commission (EC) under Project OCEANS - Overcoming challenges in the evolution and nature of massive stars, HORIZON-MSCA-2023-SE-01, No G.A 101183150 funded by the European Union. This work is part of grant CEX2025-001609-S, awarded to the Instituto de Astrofísica de Canarias under the Severo Ochoa Centre of Excellence program and funded by MICIU/AEI/10.13039/501100011033.
\end{acknowledgements}

\bibliographystyle{aa}
\bibliography{myrefs}\begin{appendix}
\onecolumn
\section{Calibrated values of the overshoot efficiency}\label{AppA}
The calibrated values of $\alpha_{\rm ov}$ in the mass range $M_{\rm ini}=12-40$\,M$_\odot$ are reported in Table \ref{table_overshoot}. They are compared to those predicted by \citet[][Eq. \ref{scott_eq}]{sco21}.
\begin{table}[H]
\centering
\caption{Calibrated $\alpha_{\rm ov}$ for stellar models in the mass range $M_{\rm ini}=12-40$\,M$_\odot$ compared to results by  \citet{sco21}. FS: calibration performed with the full sample, SR: with slow rotators only.}
\label{table_overshoot}
\begin{tabular}{cccccc}
\noalign{\hrule height 0.8pt}
\noalign{\vskip 1pt}
\noalign{\hrule height 0.8pt}
$M_{\rm ini}$ [M$_ \odot$] & Hydro models (FS) & Hydro models (SR) & Magnetic models (FS) & Magnetic models (SR) & \citet{sco21}\\
\noalign{\hrule height 0.8pt}
12 & 0.27    & 0.21 & 0.30 & 0.22 & 0.4  \\
15 & 0.35    & 0.37 & 0.39 & 0.40 & 0.49 \\
20 & 0.33    & 0.4 & 0.38 & 0.45 & 0.61 \\
25 & 0.28    & 0.35 & 0.33 & 0.40 & 0.72 \\
32 & 0.21 & 0.28 & 0.26 & 0.32 & 0.84 \\
40 & 0.18 & 0.23 & 0.20 & 0.25 & 0.94 \\
\noalign{\hrule height 0.8pt}
\end{tabular}
\end{table}
\section{Empirical velocity distribution}\label{AppB}
\begin{figure}[h]
\centering
\centerline{\includegraphics[trim=.3cm 0.3cm .3cm 0.3cm, clip=true, width=0.6\columnwidth,angle=0]{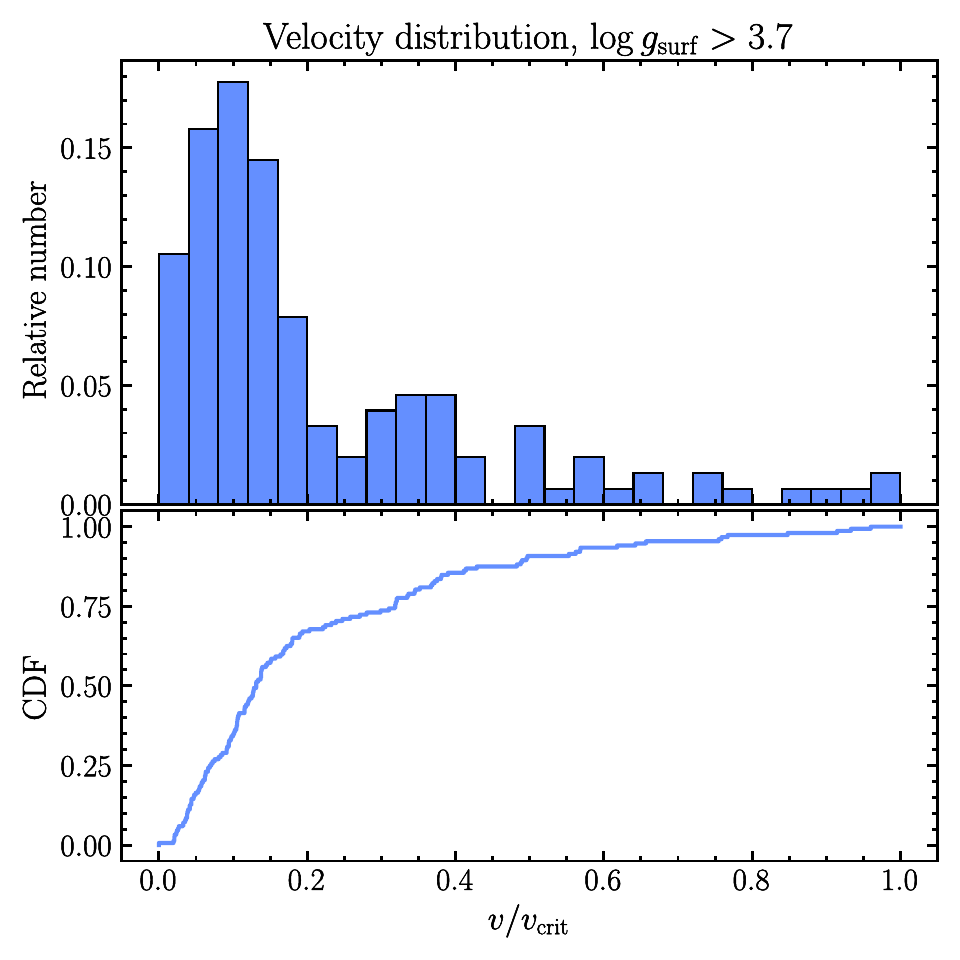}}
\caption{Empirical distribution of rotational velocities with observations from the sample by \citet{deb25}, restricted to surface gravities $\log_{\rm surf}>3.7$. \textit{Top:} Histogram. \textit{Bottom:} CDF. Observed velocities were raised by a factor $1/\braket{\sin i}=4/\pi$.}
\label{init_vel_distr}
\end{figure}

Figure \ref{init_vel_distr} shows the histogram and CDF of the empirical distribution used for the population synthesis simulations of Sect. \ref{population_comparison}. We note that the empirical distribution well catches the bimodality of the distribution. The first peak, corresponding of the bulk of the slow rotator population is at $\upsilon_{\rm ini}/\upsilon_{\rm crit}=0.1$. The second peak is found around $\upsilon_{\rm ini}/\upsilon_{\rm crit}=0.4$. The distribution shows a tail at higher rotations, in agreement with earlier studies (e.g., \citealt{hol22}). To convert from projected to intrinsic velocities, we assumed an isotropic distribution of orientations, in which case the expected value of $\sin i$ is $\braket{\sin i}=\pi/4$ \citep[see, e.g.,][]{gra88,geo14}.
\section{Synthetic noise}\label{AppC}
In Sect. \ref{population_comparison}, the synthetic population, the temperatures and luminosities were corrected accounting for gravity and limb darkening as in \citet{geo14}. The resulting $L$ and $T_{\rm eff}$ correspond to the quantities effectively measured. Then, observational uncertainties were mimicked by adding Gaussian noise to the predicted values of these quantities. The average relative uncertainties on the temperature, radius and projected velocity in the \citet{deb25} sample are respectively 
\begin{equation}
\overline{\frac{\Delta{T_{\rm eff}}}{T_{\rm eff}}}=2.4\,\%,\ \ \overline{\frac{\Delta{R}}{R}}=7\,\%, \text{ and} \ \overline{\frac{\Delta{\upsilon\sin i}}{\upsilon\sin i}}=5\,\%.
\end{equation}
For a predicted quantity $X$ (temperature, radius or projected velocity), we randomly draw a Gaussian error with $\sigma_X=X\overline{\frac{\Delta X}{X}}$ and add it to the predicted quantity. Finally, we recompute the luminosity using the Stefan-Boltzmann law $L=4\pi R^2\sigma T_{\rm eff}^4$, consistently with the approach of \citet{deb25}. This ensures a correlation between the synthetic errors on the luminosity and temperature, which is naturally present in the observed population, as a result of the approach followed by \citet{deb25}.
\section{The hot subdwarf population}\label{AppD}
\begin{figure}[h]
\centering
\centerline{\includegraphics[trim=.3cm 0.3cm .3cm 0.3cm, clip=true, width=0.8\columnwidth,angle=0]{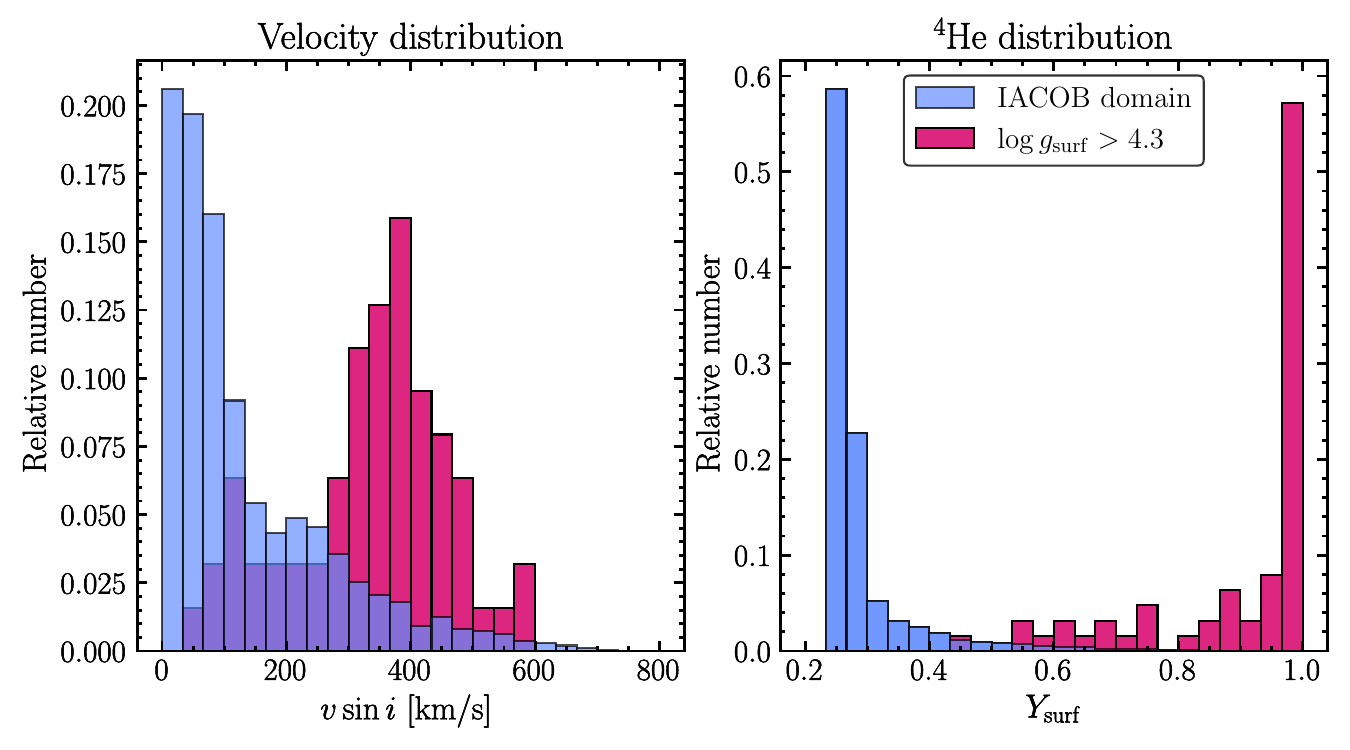}}
\caption{Predicted distributions of hot subdwarfs ($\log g_{\rm surf}>4.3$, $\log L/\text{L}_\odot=4.30-5.70$) of magnetic models with slow rotator calibration compared to the full population in the luminosity and temperature ranges of the \citet{deb25} sample ($\log L/\text{L}_\odot=4.30-5.70$, $T_{\rm eff}=14-49$\,kK). \textit{Left:} $\upsilon\sin i$ distribution. \textit{Right:} Helium mass fraction distribution.}
\label{vel_helium_subdwarfs}
\end{figure}

The sample by \citet{deb25} does not include hot subdwarfs. It is limited to objects with surface gravities $\log g_{\rm surf}<4.2$. As can be seen in Fig. \ref{HRD_syclist}, magnetic models predict a fast-rotator population to the left of the ZAMS line, due to the fact that at very fast rotations ($\upsilon_{\rm ini}/\upsilon_{\rm crit}\gtrsim 0.6$) models experience quasi-CHE. These stars are typically predicted to have large surface gravities ($\log g_{\rm surf}>4.3$) and would be observationally classified as hot subdwarfs. Given that this fast-rotator population is not predicted by single-star hydro models, observed populations of hot subdwarfs could be used in the future to constrain the rotational mixing efficiency of the models. Figure \ref{vel_helium_subdwarfs} shows the projected velocity and helium mass fraction distributions of magnetic models computed with the slow rotator calibration compared to those of the full population in the temperature and luminosity range of the \citet{deb25} sample. For the hot subdwarfs population, we also restricted the luminosity range to that of the \citet{deb25} sample.
Unlike in the full population, in which the bimodality of the velocity distribution is clearly seen, in the ``hot subdwarfs'' population the peak of the distribution is located around 400\,km/s and very few stars are found with projected velocities lower than 200\,km/s (the majority of which are attributed to low inclinations). Additionally, the majority of the stars of the full population do not show any helium enrichments, whereas in the ``hot subdwarfs'' population the large majority of the stars have $Y_{\rm surf}\gtrsim 0.8$, consistently with them undergoing quasi-CHE caused by rapid rotation. These results are key predictions which could be tested in the future against populations of hot subdwarfs. Interestingly, the rotational mixing efficiency of hydro and magnetic models was calibrated with the same observable constraints (see Appendix A in \citealt{sci26}), at lower initial velocities. The fact that only magnetic models experience quasi-CHE indicates that the mixing velocity dependence is very pronounced in this case. It is important to note that if most fast rotators are indeed binary interaction products, their chemical evolution likely differs from what standard single-star evolution predicts. If mass transfer occurs after a significant composition gradient has already established, it is not guaranteed that rotational mixing will be strong enough for making the stars experience quasi-CHE \citep[see, e.g.,][]{mey97}. In this case, the absence of a population of fast-rotating, helium enriched hot subdwarfs does not necessarily rule out the magnetic models' rotational mixing efficiency.
\end{appendix}
\end{document}